\documentclass[12pt]{iopart}

\usepackage[T1]{fontenc}
\usepackage[british]{babel}
\usepackage{amssymb}
\usepackage{cite}
\usepackage{iopams}
\usepackage{color}
\usepackage{mathrsfs}
\usepackage{graphicx}

\begin{document}

\title[Fractional Brownian motion in superharmonic potentials]{Fractional Brownian
motion in superharmonic potentials and non-Boltzmann stationary distributions}

\author{Tobias Guggenberger$^\dagger$, Aleksei Chechkin$^{\dagger,\ddagger}$ and
Ralf Metzler$^\dagger$}

\address{$^\dagger$ Institute of Physics \& Astronomy, University of Potsdam,
14476 Potsdam-Golm, Germany\\
$^\ddagger$ Akhiezer Institute for Theoretical Physics, Kharkov 61108, Ukraine}
\eads{\mailto{rmetzler@uni-potsdam.de} (Corresponding author: Ralf Metzler)}

\begin{abstract}
We study the stochastic motion of particles driven by long-range correlated
fractional Gaussian noise in a superharmonic external potential of the form
$U(x)\propto x^{2n}$ ($n\in\mathbb{N}$). When the noise is considered to be
external, the resulting overdamped motion is described by the non-Markovian
Langevin equation for fractional Brownian motion. For this case we show the
existence of long time, stationary probability density functions (PDFs) the
shape of which strongly deviates from the naively expected Boltzmann PDF in
the confining potential $U(x)$.  We analyse in detail the temporal approach 
to stationarity as well as the shape of the non-Boltzmann stationary PDF. A
typical characteristic is that subdiffusive,  antipersistent (with negative
autocorrelation) motion tends to effect an accumulation of probability close
to the origin as compared to the corresponding Boltzmann distribution while
the opposite trend occurs for superdiffusive (persistent) motion.  For this
latter case this leads to distinct bimodal shapes of the PDF. This property
is compared to a similar phenomenon observed for Markovian L{\'e}vy flights
in superharmonic potentials. We also demonstrate that the motion encoded in
the fractional Langevin equation driven by fractional Gaussian noise always
relaxes to the Boltzmann distribution, as in this case the
fluctuation-dissipation theorem is fulfilled.
\end{abstract}

\section{Introduction}
\label{sec:introduction}

Robert Brown's vivid account on the "rapid oscillatory motion" of what he
called "Molecules" \cite{brown}, micron-sized granules contained in the
pollen grains of the flowering plant \emph{Clarckia pulchella}\footnote{Now
renamed \emph{Clarkia pulchella}.}, inspired the probabilistic theoretical
formulations of diffusive processes by Einstein \cite{einstein},
Sutherland \cite{sutherland}, and Smoluchowski \cite{smoluchowski}. A
central idea of stochastic processes \cite{vankampen} was the concept
of fluctuating forces \cite{brenig,zwanzig} originally formulated by
Langevin in his description of "Brownian motion" based on the extension of
Newton's second law \cite{langevin,coffey}. Brownian motion as a mathematical
model \cite{levy} soon emerged
as a cornerstone of non-equilibrium statistical physics and physical
kinetics \cite{zwanzig,landau}.  Notably, the theoretical predictions of
Einstein, Sutherland, Smoluchowski, and Langevin were soon confirmed by
experimentalists such as Perrin \cite{perrin}, Nordlund \cite{nordlund},
and Kappler \cite{kappler}.

While normal Brownian diffusion is characterised by the linear time dependence
of the mean squared displacement (MSD) $\langle X^2(t)\rangle=\int_{-\infty}^{
\infty}x^2P(x,t)dx$, anomalous diffusion has the non-linear form $\langle X^2(
t)\rangle\simeq K_{\alpha}t^{\alpha}$, with the anomalous diffusion exponent
$\alpha$ and the generalised diffusion coefficient $K_{\alpha}$ of physical
dimension $\mathrm{cm}^2/\mathrm{sec}^{\alpha}$. Depending on the value of
$\alpha$ we distinguish subdiffusion ($0<\alpha<1$) and superdiffusion
($\alpha>1$) \cite{bouchaud1990,metzler2000}. A wide range
of observations of anomalous diffusion come from biological and soft matter
systems \cite{norregaard2017,hoefling2013,pt}. Subdiffusion was reported for
tracer motion in living cells \cite{caspi,jeon2011,weiss2004,bronstein2009,
burnecki2012,weber2010,tabei2013} and in complex liquids \cite{jeon2013,
szymanski,wong2004,yael,banks2005}. Superdiffusion, based on active motion
in living cells, was studied in \cite{caspi,seisenhuber,reverey2015,jae_neuro,
samu_jcp,granick_lw}. We also mention results for anomalous diffusion
from supercomputing and experimental studies of membrane systems
\cite{jeon2012,jeon2016,kneller,gerald,natcom,weigel2011,weigel2013},
internal protein dynamics \cite{jeremy,xie}, or the motion along
membranes and other surfaces \cite{liang,amanda,yamamoto1,krapf2016,pt1}.

As suggested by Einstein \cite{einstein}, normal diffusion with its
linear time dependence of the MSD and the Gaussian probability density function
(PDF) can be understood as resulting from a random motion consisting of a
sequence of random displacements which satisfy the following three conditions:
(i) There is a finite correlation time after which individual displacements
become stochastically independent, (ii) the displacements are identically
distributed, and (iii) the displacements have a finite second moment.
Anomalous diffusion may appear whenever one of these conditions is
violated \cite{bouchaud1990,metzler2014,metzler2000}. We here consider the
case when long-range temporal correlations of the displacements effect
anomalous diffusion. The prototypical example for this
mechanism is the Mandelbrot-van Ness fractional Brownian motion (FBM), a
Gaussian, non-Markovian process with stationary increments
\cite{mandelbrot1968,qian2003}.
FBM has been identified to give rise to the anomalous-diffusive
behaviour in a variety of systems including tracer motion in complex liquids
\cite{jeon2013,szymanski2009}, in living cells \cite{weber2010,tabei2013,
magdziarz2009,jeon2011,krapf2019} and in membrane dynamics \cite{jeon2012,jeon2016,
kneller}.
FBM-like correlations are particularly studied in modern financial market models
to account for market "roughness" \cite{comte1998,rostek2013,euch}, and similar
effects in network traffic \cite{mikosch2002}. FBM was also applied to describe
observed density profiles of serotonergic brain fibres \cite{janusonis2020}.

For many experimentally relevant scenarios it is important to
study the behaviour of a stochastic motion under confinement. When this motion
is normal Brownian diffusion, at long times the equilibrium state is characterised
by the Boltzmann distribution, a fundamental property of statistical mechanics
\cite{landau1}. Thus, for a particle in a confining external potential $V(x)$,
in the overdamped limit $\lim_{t\to\infty}P(x,t)=\mathscr{N}\exp(-\beta V(x))$,
where $\beta=1/[k_BT]$ is the Boltzmann factor and $\mathscr{N}$
a normalisation. Typically, the dynamical
approach to this distribution from a non-equilibrium state is described by the
Fokker-Planck-Smoluchowski equation \cite{risken}. Confinement can be induced
when the tracer is measured in an optical tweezers setup, exerting a Hookean
restoring force on the tracer \cite{jeon2013,jeon2011}. In internal protein
dynamics relative motion of two aminoacids is effectively harmonically confined 
through the protein backbone connecting these aminoacids \cite{xie,jeremy},
similar to the confinement of a tracer induced by a linker molecule \cite{simon}.
Inside the cell, tracers are confined by the cell walls or internal membrane
barriers. Growing serotonergic brain fibres are confined in specific brain
regions \cite{janusonis2020}.

To study confinement effects for systems driven by
fractional Gaussian noise (FGN), one of the main frameworks
is based on the Langevin equation for FBM and the fractional Langevin equation
(FLE) introduced below. If we assume the validity of the fluctuation-dissipation
theorem (FDT) \cite{kubo1966}, the system equilibrates to the
stationary Boltzmann distribution and has a well-defined temperature. In this
case, the driving FGN is called "internal" and represents the "bath" variables
\cite{zwanzig}. However, especially in biological systems the equilibrium
assumption fails due to perpetual, energy-consuming active processes.
Following Klimontovich the measured anomalous diffusion can be described
by the Langevin equation driven by "external" noise \cite{klimontovich1995},
which we here model by FGN. Recent numerical studies of the
overdamped Langevin equation with reflecting boundary conditions and driven by
FGN showed that the long-time stationary PDF deviates strikingly from the
Boltzmann distribution, which in this case would be a uniform distribution
\cite{guggenberger2019,vojta2020,janusonis2020}. In fact a similar situation
is known for Markovian L{\'e}vy flights in a superharmonic potential $U(x)
\propto x^{2n}$, $n\in\mathbb{N}$, which can be described by an overdamped
Langevin equation driven by L{\'e}vy stable noise. It was shown that the
long-time stationary PDF disagrees with the Boltzmann distribution even for
the harmonic case \cite{sune,chech} and even features distinct bimodal and transient
multimodal shapes despite the unimodal confining potential \cite{chechkin2002,
chechkin2003,chechkin2004,karol}.
For the FLE the PDF for the free (no confining potential) and harmonic potential
cases were obtained \cite{kou2008,metzler2012}.
The FLE has also been studied with reflecting boundary
conditions and a flat potential with exponential confining "wings"
\cite{metzler2010,holmes2019,vojta2019}.

We here investigate numerically the diffusive motion of particles confined in
a superharmonic external potential, based on the overdamped Langevin equation
driven by FGN. We find that the stationary PDF deviates significantly from the
naively expected Boltzmann distribution. Remarkably, the stationary PDF exhibits
a distinct bimodal shape for steeper than harmonic potentials, despite the
unimodal nature of the confining potential. The resemblance with results found
for L{\'e}vy flights in superharmonic potential \cite{chechkin2002,chechkin2003,
chechkin2004,karol} is discussed. Finally we demonstrate that for the overdamped
FLE the PDF converges to the Boltzmann distribution, as it should.

\section{Fractional Brownian motion and fractional Gaussian noise}
\label{sec:fgnAndFbm}

Fractional Brownian motion (FBM) is a centred Gaussian process with
two-time auto-covariance function
\begin{equation}
\label{eq:autocovFbm}
\langle X(t_1)X(t_2)\rangle=K_{\alpha}\big[t_1^\alpha+t_2^\alpha-|t_1-t_2|^\alpha\big],
\,\,\,0<\alpha\le2,
\end{equation}
including the MSD $\langle X^2(t)\rangle=2K_{\alpha}t^\alpha$ for $t_1=t_2=t$.
For $\alpha=1$ the correlations
vanish, and FBM reduces to Brownian motion. The PDF of FBM for natural boundary
conditions ($\lim_{|x|\to\infty}P(x,t)=0$) is the Gaussian
\begin{equation}
\label{eq:pdfFbm}
P(x,t)=\frac{1}{\sqrt{4\pi K_{\alpha}t^\alpha}}\exp\left(-\frac{x^2}{4K_{\alpha}
t^\alpha}\right).
\end{equation}
Since the sample paths of FBM are almost surely continuous but not differentiable
\cite{mandelbrot1968}, we follow Mandelbrot and van Ness and define FGN as the
difference quotient \cite{mandelbrot1968}
\begin{equation}
\label{eq:fgnDef}
\xi(t;\delta t)=\frac{X(t+\delta t)-X(t)}{\delta t},
\end{equation}
where $\delta t$ is a small but finite time step. It follows that FGN is a
centred stationary Gaussian process whose auto-covariance function is readily
obtained from \eref{eq:autocovFbm} and \eref{eq:fgnDef},
\begin{equation}
\label{eq:fgnCov}
\langle\xi(t;\delta t)\xi(t+\tau;\delta t)\rangle=K_\alpha (\delta t)^{\alpha-2}
\left[\left|\frac{\tau}{\delta t}+1\right|^\alpha+\left|\frac{\tau}{\delta t}-1
\right|^\alpha-2\left|\frac{\tau}{\delta t}\right|^\alpha\right].
\end{equation}
The variance of FGN is thus $\langle\xi^2(t;\delta t)\rangle=2K_\alpha(\delta t)
^{\alpha-2}$. At times much longer than the time step, $\tau\gg\delta t$, one
has \footnote{Using $(x\pm1)^\alpha=\sum_{k=0}^\infty\frac{1}{k!}\frac{d^k
x^\alpha}{dx^k}(\pm1)^k$.}
\begin{equation}
\label{eq:fgnCovAsy}
\langle\xi(t;\delta t)\xi(t+\tau;\delta t)\rangle\sim K_\alpha\alpha(\alpha-1)
\tau^{\alpha-2},
\end{equation}
and hence the correlations are positive (negative) for $\alpha>1$ ($\alpha<1$).
Considering $\delta t$ to be "infinitesimally small" we write $\xi(t)$ and
formally take it as the "derivative" of FBM with covariance \eref{eq:fgnCovAsy}
($\tau>0$), so that $X(t)=\int_0^t\xi(t')dt'$.
Finally we mention that
\begin{eqnarray}
\label{eq:fgnCovInt}
\nonumber
\fl\int_0^\infty\langle\xi(t;\delta t)\xi(t+\tau;\delta t)\rangle d\tau&=&K_\alpha
(\delta t)^{\alpha-1}\left[\int_0^1[(s+1)^{\alpha}+(1-s)^{\alpha}-2s^{\alpha}]ds
\right.\\
\nonumber
\fl
&&\left.+\int_1^\infty[(s+1)^{\alpha}+(s-1)^{\alpha}-2s^{\alpha}]ds\right]\\
\nonumber
\fl
&=&\frac{K_\alpha (\delta t)^{\alpha-1}}{\alpha+1}\lim_{s\to\infty}[(s+1)^{
\alpha+1}+(s-1)^{\alpha+1}-2s^{\alpha+1}]\\
\fl
&=&\left\{\begin{array}{ll}0,&0<\alpha<1\\K_\alpha,&\alpha=1\\\infty,&1<\alpha<2
\end{array}\right..
\end{eqnarray}
Equations \eref{eq:fgnCovAsy} and \eref{eq:fgnCovInt} demonstrate a drastic
difference between persistent ($1<\alpha<2$) and antipersistent ($0<\alpha
<1$) FGNs, in particular, note the vanishing integral over the noise
auto-covariance.

\subsection{Overdamped Langevin equation for FBM}
\label{sec:langevinEq}

The overdamped Langevin equation for a test particle performing FBM reads
\begin{equation}
\label{eq:langevin}
\frac{dX(t)}{dt}=-\frac{1}{m\zeta}\frac{dU}{dx}(X(t))+\xi(t),\,\,\,U(x)=
\frac{k}{2n} x^{2n},
\end{equation}
where $X(t)$ is the particle position and $U(x)$ is the external potential,
where $k>0$ and $n\in\mathbb{N}$. This potential is harmonic for $n=1$ and
will be called "superharmonic" for $n>1$. The particle is driven by the
FGN $\xi(t)$, $m$ is the mass of the particle and $\zeta$ is a friction
coefficient of dimension $1/\mathrm{sec}$. The Langevin equation
\eref{eq:langevin} can be rewritten as
\begin{equation}
\label{eq:normLangevin}
\frac{dX(t)}{dt}=-\frac{dU}{dx}(X(t))+\eta(t),\,\,\,U(x)=\frac{x^{2n}}{2n}
\end{equation}
in dimensionless form, where $\eta(t)=\xi(t)/\sqrt{2K_{\alpha}}$ is a
normalised FGN. The two relevant parameters defining
our system are the anomalous diffusion exponent $\alpha$ and the power
exponent $n$ of the potential.

\begin{figure}
\includegraphics[width=0.49\textwidth]{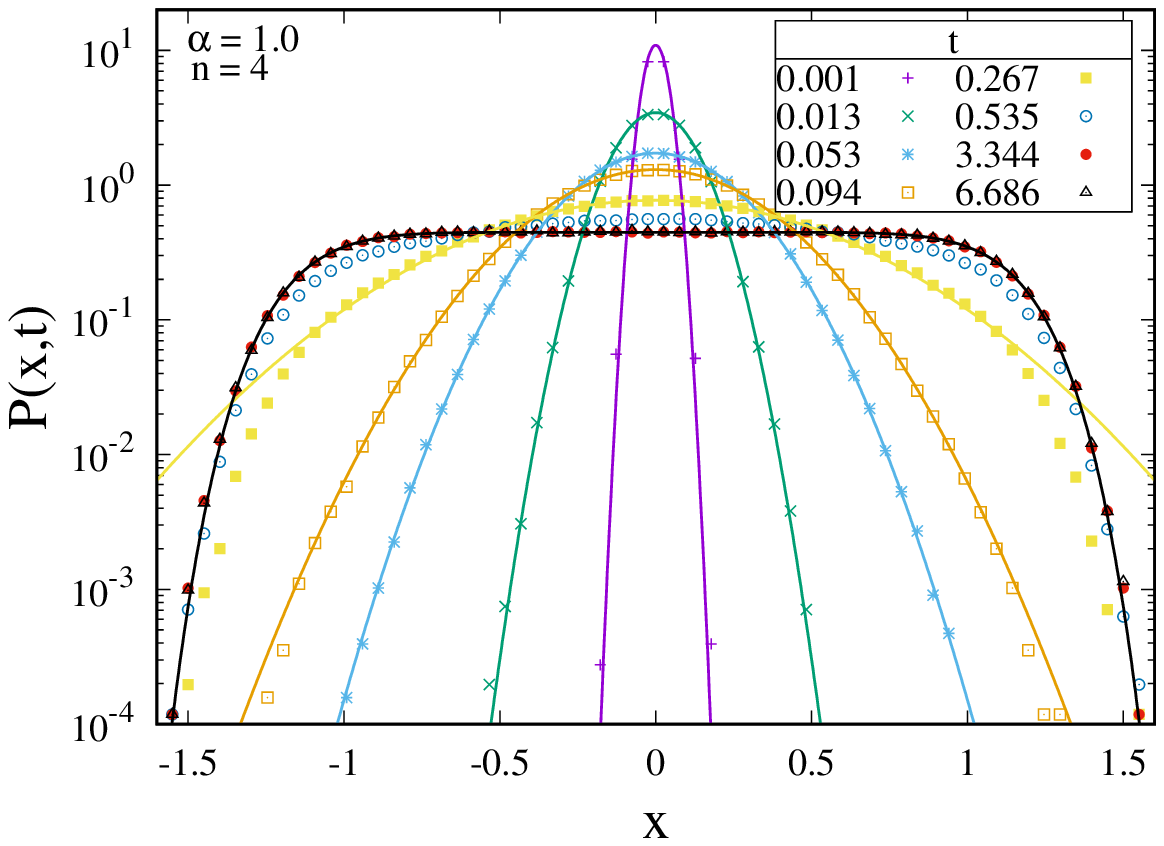}
\includegraphics[width=0.49\textwidth]{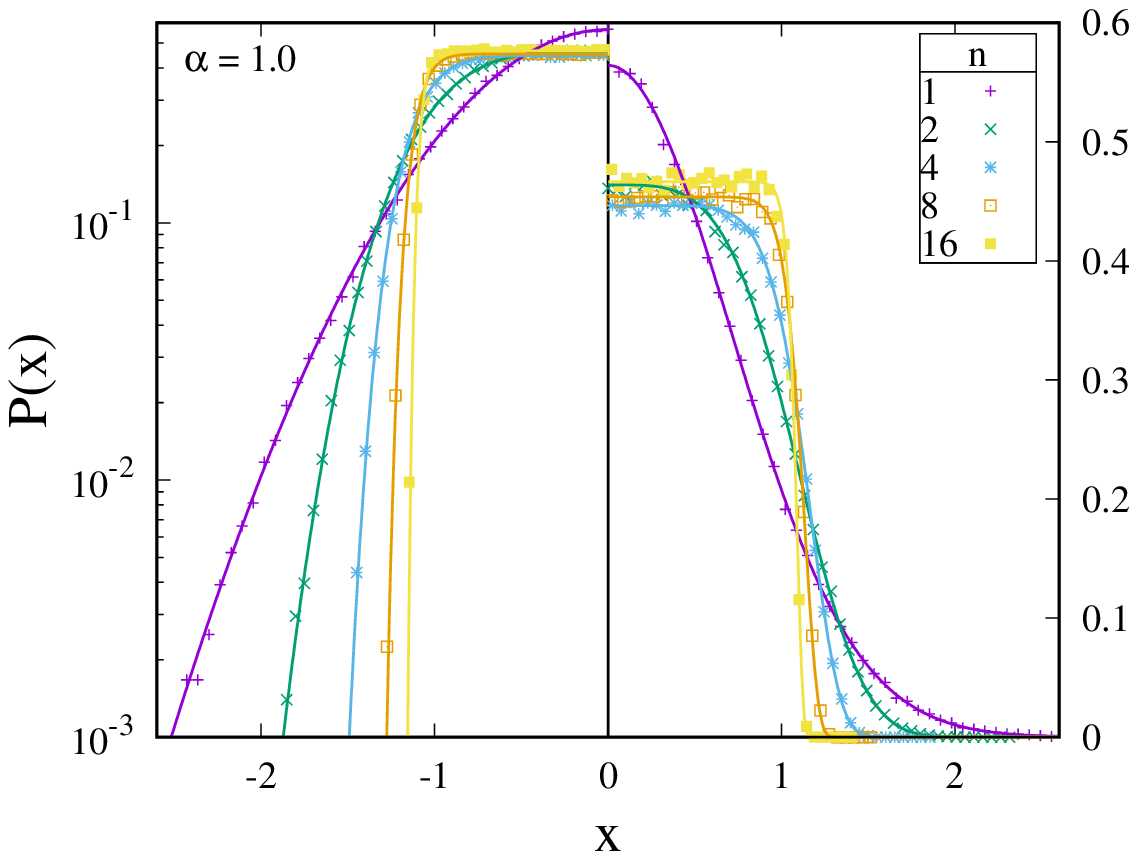}
\caption{PDF for uncorrelated white Gaussian noise ($\alpha=1$). Left: time
evolution of the PDF for a superharmonic potential with $n=4$. The black line
shows the theoretical Boltzmann distribution \eref{eq:solStatFpeBrownian2},
the coloured lines show the (rescaled) Gaussian \eref{eq:pdfFbm} for free
motion. Note the increasing deviation of the simulations from this short-time
form, as it should be. Note that the "strange" values for the time $t$ arise
due to the rescaling to dimensionless variables. Right: stationary PDF for
different superharmonic potentials. The coloured lines show the Boltzmann
distribution \eref{eq:solStatFpeBrownian2}. For $x<0$ the (symmetric) data
is plotted logarithmically (left axis) and for $x\geq0$ linearly (right axis)}
\label{fig:pdfUncorrNoise}
\end{figure}

We note that the FGN in the Langevin equations (\ref{eq:langevin}) and
(\ref{eq:normLangevin}) is external, i.e., the FDT is not fulfilled for
$\alpha\neq1$. Thus while for any process driven by a random force with
finite varience a confining potential will lead to the emergence of a
stationary PDF at long times, this will not be
a thermalised Boltzmann distribution, see below. Only in the uncorrelated
case ($\alpha=1$), when the FGN is a white
Gaussian unit noise, $\langle\eta(t_1)\eta(t_2)\rangle=\delta(t_1-t_2)$ and
hence the PDF $P(x,t)$ satisfies the Fokker-Planck-equation, the stationary
state is given by the Boltzmann distribution. For the initial
condition $P(x,0)=\delta(x)$ it is straightforward to show that the stationary
solution $P(x)=\lim_{t\to\infty}P(x,t)$ is the Boltzmann distribution
\begin{equation}
\label{eq:solStatFpeBrownian2}
P(x)=\mathscr{N}_1(n)\exp\left(-\frac{x^{2n}}{n}\right),\,\,\,\mathscr{N}_1(n)=
\frac{n^{1-1/(2n)}}{\Gamma\left(1/(2n)\right)}
\end{equation}
where $\Gamma(z)=\int_0^\infty s^{z-1}e^{-s}ds$ is the Gamma function. Figure
\ref{fig:pdfUncorrNoise} shows simulations results for the time-dependent
and stationary PDF for the Brownian case $\alpha=1$. At short times the PDF
is given by the (rescaled) Gaussian
\eref{eq:pdfFbm} for the free case, and it relaxes towards the Boltzmann PDF
\eref{eq:solStatFpeBrownian2}.

\section{Results}
\label{sec:results}

\subsection{Harmonic potential}
\label{sec:LEqHarPot}

For the harmonic potential ($n=1$) the PDF $P(x,t)$ for FBM satisfies the
Fokker-Planck equation
\cite{adelmann}
\begin{equation}
\label{eq:fpeHarmonic}
\frac{\partial}{\partial t}P(x,t)=\frac{\partial}{\partial x}\left[xP(x,t)\right]
+D(t)\frac{\partial^2}{\partial x^2}P(x,t),
\end{equation}
with initial condition $P(x,0)=\delta(x)$ and the time-dependent coefficient
\begin{equation}
\label{eq:fpeHarmonic2}
\fl D(t)=\int_0^te^{-\tau}\langle\eta(t)\eta(t+\tau)\rangle d\tau
=\frac{\alpha}{2}\left(t^{\alpha-1}e^{-t}+\gamma(\alpha,t)\right)\stackrel{t\to
\infty}{\longrightarrow}\frac{\Gamma(\alpha+1)}{2},
\end{equation}
where $\gamma(z,t)=\int_0^ts^{z-1}e^{-s}ds$ is the incomplete gamma function.
The PDF has the Gaussian form and long time limit
\begin{equation}
\label{eq:solStatFpeHarmonic}
\fl P(x,t)=\frac{1}{\sqrt{2\pi\langle X^2(t)\rangle}}\exp\left(-\frac{x^2}{2\langle
X^2(t)\rangle}\right)\stackrel{t\to\infty}{\longrightarrow}\frac{1}{\sqrt{\pi\Gamma
(\alpha+1)}}\exp\left(-\frac{x^2}{\Gamma(\alpha+1)}\right).
\end{equation}
The MSD follows directly from the Langevin equation \cite{metzler2012},
\begin{equation}
\label{eq:secMomLeqHarPot}
\fl\langle X^2(t)\rangle=t^\alpha e^{-t}+\frac{\gamma(\alpha+1,t)}{2}-\frac{t^{
\alpha+1}e^{-2t}}{2(\alpha+1)}M(\alpha+1,\alpha+2,t)\stackrel{t\to\infty}{
\longrightarrow}\frac{\Gamma(\alpha+1)}{2},
\end{equation}
in terms of the Kummer function $M(a,b,z)$ \cite{abramowitz72}. The explicit
$\alpha$-dependence in the stationary solution
(\ref{eq:solStatFpeHarmonic}) and the corresponding stationary MSD
(\ref{eq:secMomLeqHarPot}) demonstrates the non-thermal character of FBM.

\subsection{Superharmonic potential}
\label{sec:LEqSupHarPot}

As we have seen in the case of uncorrelated noise ($\alpha=1$) as well as in the
case of an harmonic potential ($n=1$), the PDF converges at long times to the
Boltzmannian
\begin{equation}
\label{eq:boltzmannForm}
P(x)=\mathscr{N}_2\exp\Big(-aU(x)\Big),
\end{equation}
with $a > 0$ and normalisation constant $\mathscr{N}_2$.\footnote{For uncorrelated
noise we have $a=2$ and for correlated noise in the harmonic potential we have
$a=2/\Gamma(\alpha+1)$.}
One may expect that in the case of correlated Gaussian noise with a superharmonic
potential ($\alpha\neq1$, $n\geq2$) the PDF still relaxes to the Boltzmannian,
however, we will show that the actual form of the stationary PDF strongly deviates
from the form \eref{eq:boltzmannForm}.

\begin{figure}
\centering
\includegraphics[width=0.49\textwidth]{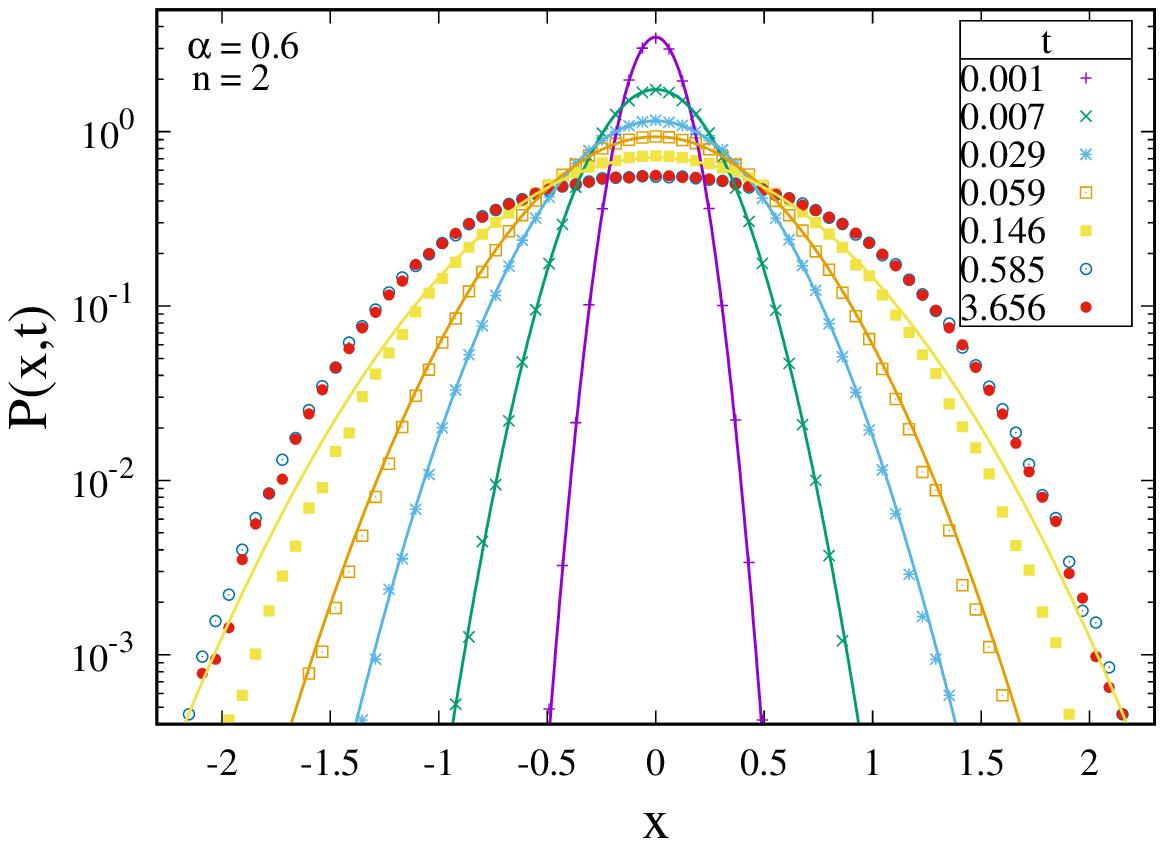}
\includegraphics[width=0.49\textwidth]{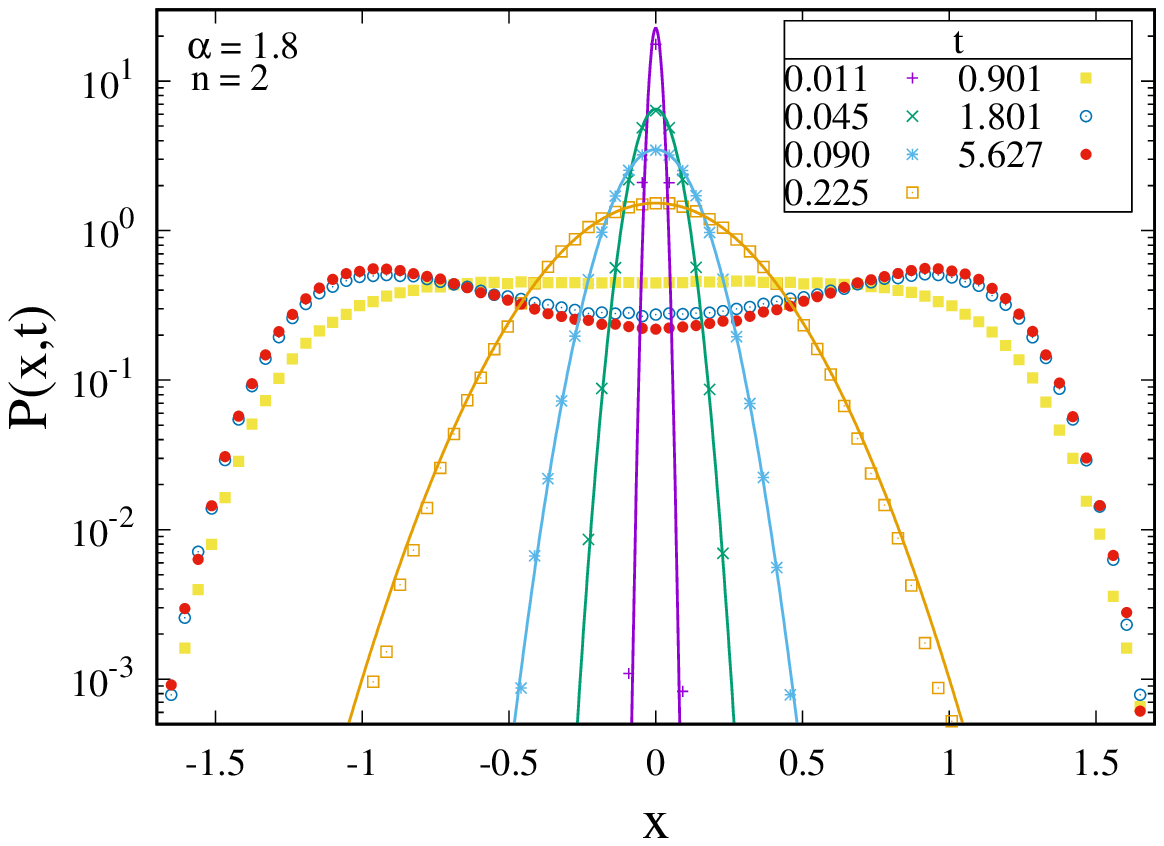}
\includegraphics[width=0.49\textwidth]{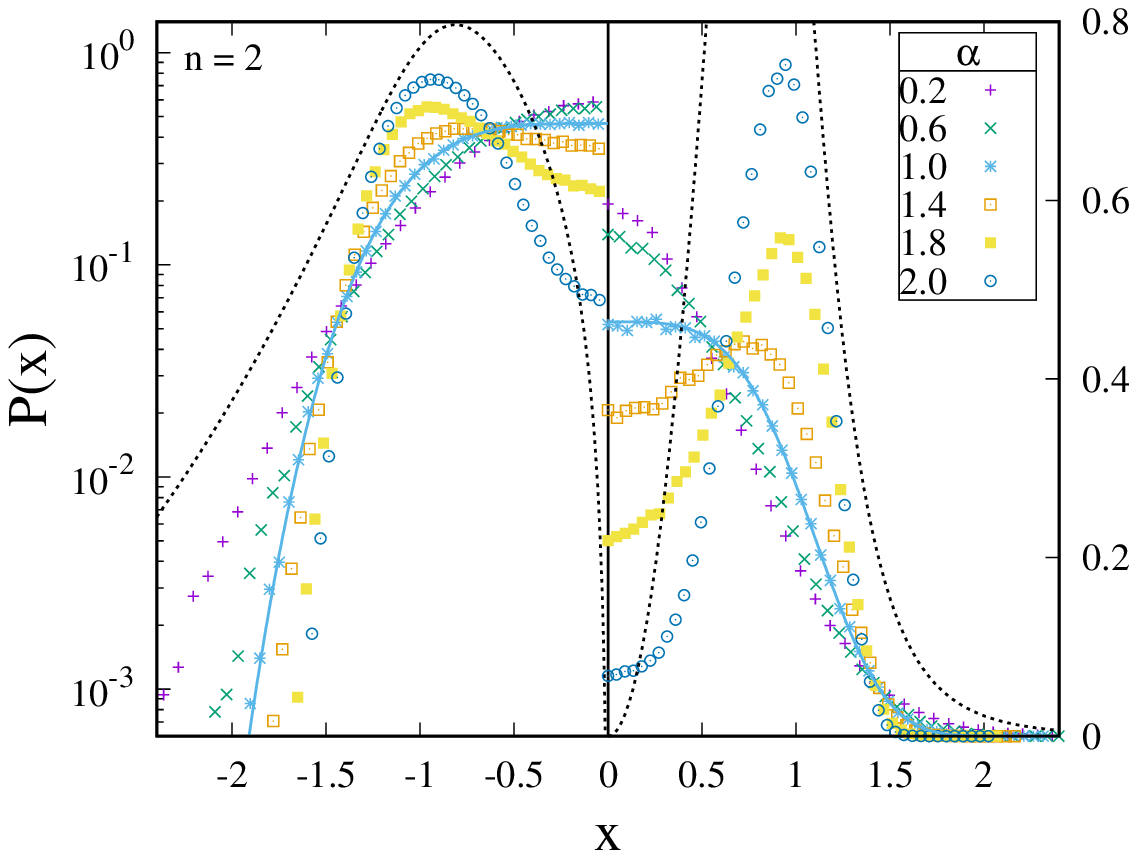}
\caption{PDF for superharmonic potential with $n=2$. Top: time dependence of
PDFs for negatively ($\alpha=0.6$, left) and positively ($\alpha=1.8$, right)
correlated FGN.  The coloured lines show the corresponding (rescaled) Gaussian
\eref{eq:pdfFbm} of the free case. Bottom: Stationary PDF for different values
of $\alpha$. The solid blue line represents the theoretical stationary PDF
\eref{eq:solStatFpeBrownian2} for uncorrelated white Gaussian noise ($\alpha=1$).
The dashed line shows the curvature \eref{eq:curvPot} of the potential. For $x<
0$ the data is plotted logarithmically (left axis) and for $x\geq0$ linearly
(right axis). For details of the numerical scheme see \ref{sec:numerics}.}
\label{fig:pdfCorrNoiseSupHarmPot}
\end{figure}

Figure \ref{fig:pdfCorrNoiseSupHarmPot} shows simulation results for the
time-dependent and stationary PDF in a quartic potential ($n=2$). Note that in
the ballistic limit $\alpha=2$, for which the FGN becomes time-independent and
perfectly correlated, one has $\int_0^t\eta(\tau)d\tau=Zt$, where the amplitude $Z$
is distributed like a unit Gaussian, $Z\stackrel{\mathrm d}{=}\mathcal{N}(0,1)$.
Hence, in the ballistic limit, the randomness of the motion is solely due to a
random "initial velocity" $Z$.

At short times the PDF agrees with the (rescaled) Gaussian \eref{eq:pdfFbm} in
the free (unconfined) case. At long times the PDF converges to a stationary PDF
that is significantly different from the Boltzmannian \eref{eq:boltzmannForm}.
For positively correlated FGN ($\alpha>1$) this difference is quite conspicuous,
as the stationary PDF exhibits a distinct bimodal shape despite the unimodal
and symmetric shape of the confining potential with a unique global minimum at
the origin. Indeed, the stationary PDF has two global maxima located symmetrically
with respect to the origin, and a local minimum at the origin. The amplitude of
the two peaks increases with $\alpha$, and the plot shows that their position is
close to, but not exactly at the maxima of the curvature of the potential,
\begin{equation}
\label{eq:curvPot}
r(x)=\frac{\left|U''(x)\right|}{\Big(1+(U'(x))^2\Big)^{3/2}}=\frac{(2n-1)
x^{2n-2}}{\Big(1+x^{4n-2}\Big)^{3/2}}.
\end{equation}
For negatively correlated FGN ($\alpha<1$) the difference of the stationary PDF
to the Boltzmannian \eref{eq:boltzmannForm} is
less obvious, as the PDF is still unimodal. We note that
in figure \ref{fig:pdfCorrNoiseSupHarmPot} for $n=2$ the tails of the stationary
PDF decay progressively faster with increasing anomalous diffusion exponent
$\alpha$. This is in contrast to the harmonic case ($n=1$) in which the tails
of the PDF decay slower for increasing $\alpha$, due to the dependence of the
width  $a=2/\Gamma(\alpha+1)$ in the corresponding Boltzmannian
\eref{eq:boltzmannForm}.

To analyse what controls the shape of the tails we computed the kurtosis $\kappa$
of the simulated stationary PDFs that is defined as the fourth standardised moment
$\kappa=\langle(X-\langle X\rangle)^4\rangle/\langle(X-\langle X\rangle)^2
\rangle^2$ and compare it to the kurtosis of the Boltzmannian
\eref{eq:boltzmannForm},
\begin{equation}
\label{eq:curtSuperGauss}
\kappa=\Gamma\left(\frac{5}{2n}\right)\Gamma\left(\frac{1}{2n}\right)\Big/
\Gamma^2\left(\frac{3}{2n}\right).
\end{equation}
For the harmonic potential ($n=1$) the Boltzmannian is Gaussian with
$\kappa =3$ while for the quartic potential ($n=2$) the Boltzmannian is
"super-Gaussian" with $\kappa\approx2.19$. This platykurtic value means that
the distribution falls off faster than a Gaussian normal distribution and
thus has fewer and less extreme outliers. The crucial observation is that
the kurtosis of the Boltzmannian only depends on $n$ and not on $a$. Table
\ref{tab:kurtosis} lists the kurtosis of the stationary PDFs for the quartic
potential ($n=2$) shown in figure \ref{fig:pdfCorrNoiseSupHarmPot}.  As can be
seen, the kurtosis monotonically decreases with $\alpha$. For the uncorrelated
case ($\alpha=1$) the kurtosis of the simulated data exactly reproduces
the theoretical value $\kappa\approx2.19$. For positively correlated FGN
($\alpha > 1$) the kurtosis values are smaller than this value, while for
negatively correlated FGN ($\alpha < 1$) they are larger, in agreement with
the observed faster or slower decay of the tails. We thus conclude that also
for negatively correlated FGN ($\alpha<1$) the stationary PDF is fundamentally
different to the Boltzmannian \eref{eq:boltzmannForm}.

\begin{table}
\centering
\begin{tabular}{ccc}
\hline\hline
$\alpha$ & $\kappa$ & $b$ \\ \hline
0.2 & 2.94 & 0.52\\
0.6 & 2.55 & 0.67\\
1.0 & 2.19 & 1.00\\
1.4 & 1.88 & \\
1.8 & 1.56 & \\\hline\hline
\end{tabular}
\caption{Kurtosis $\kappa$ of the simulated stationary PDFs in the quartic
potential ($n=2$), figure \ref{fig:pdfCorrNoiseSupHarmPot}, for different
anomalous diffusion exponents $\alpha$. For uncorrelated noise ($\alpha=1$)
the value $2.19$ expected for the Boltzmannian \eref{eq:boltzmannForm} is
nicely reproduced. We also show the stretching exponent
$b$ (see equation (\ref{eq:supGauPdf})) for the unimodal subdiffusive
cases, as extracted from the kurtosis $\kappa$. For the Brownian case
($\alpha=1$) the expected value $b=1$ is consistently recovered in this
procedure.}
\label{tab:kurtosis}
\end{table}

\begin{figure}
\centering
\includegraphics[width=0.49\textwidth]{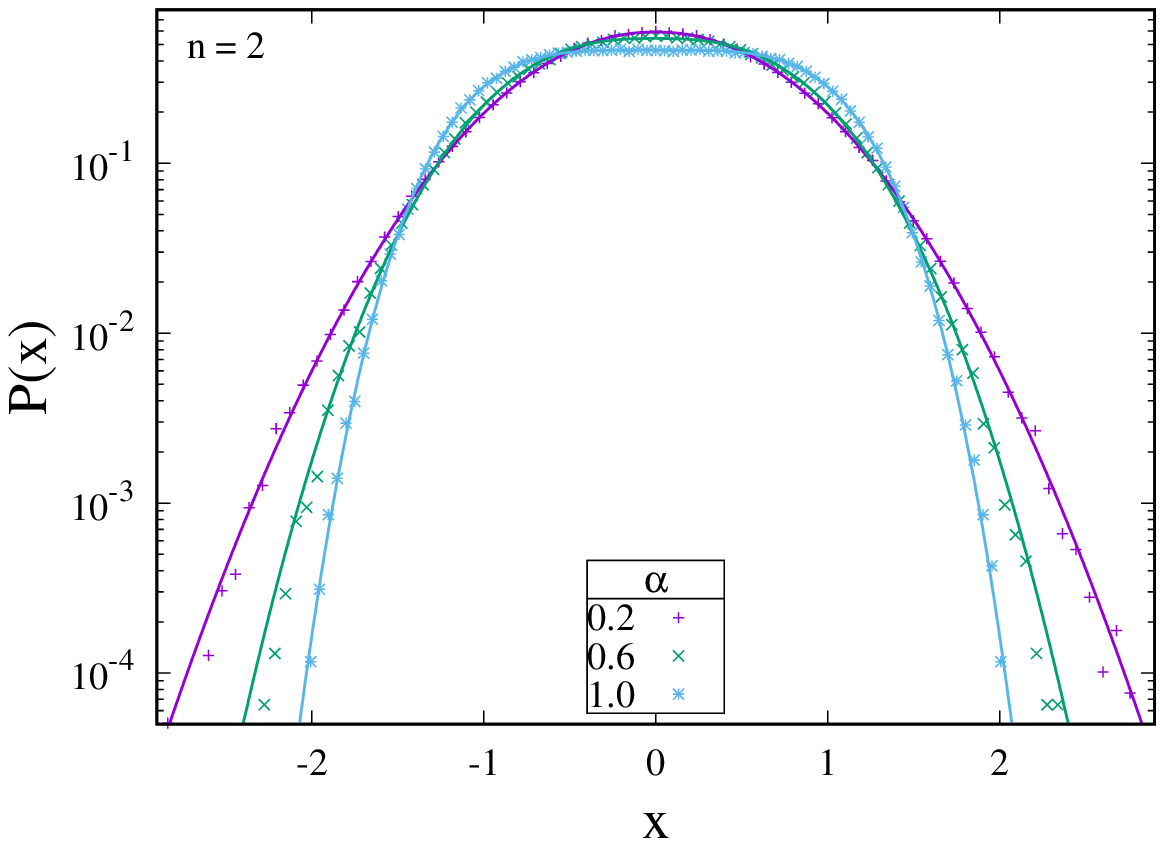}
\includegraphics[width=0.5\textwidth]{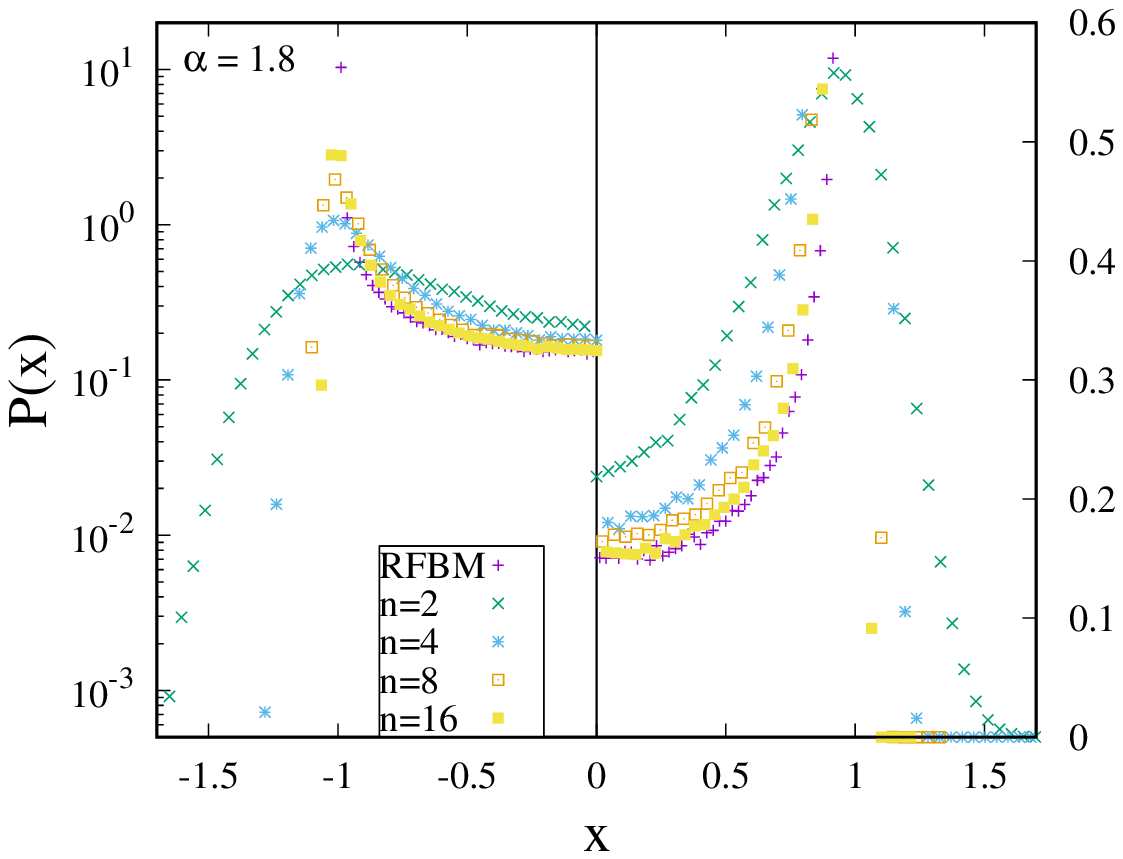}\\
\includegraphics[width=0.49\textwidth]{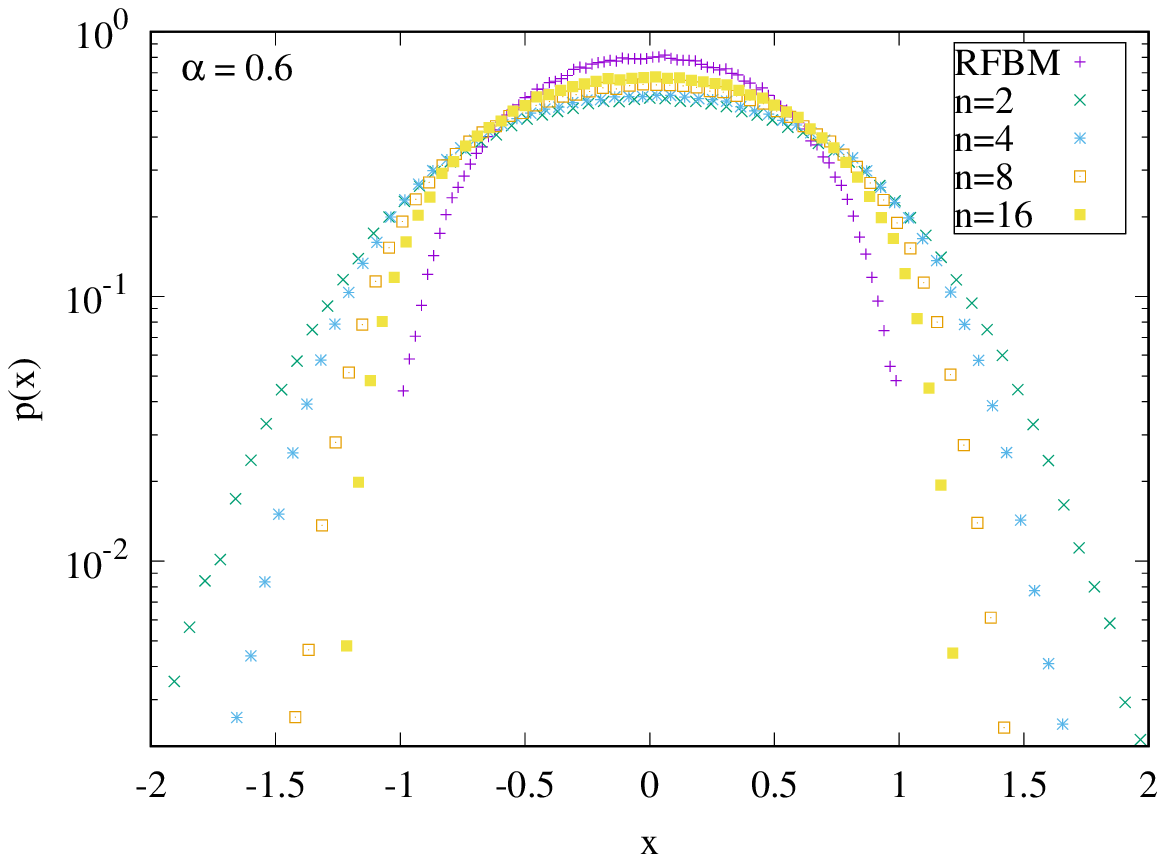}
\includegraphics[width=0.49\textwidth]{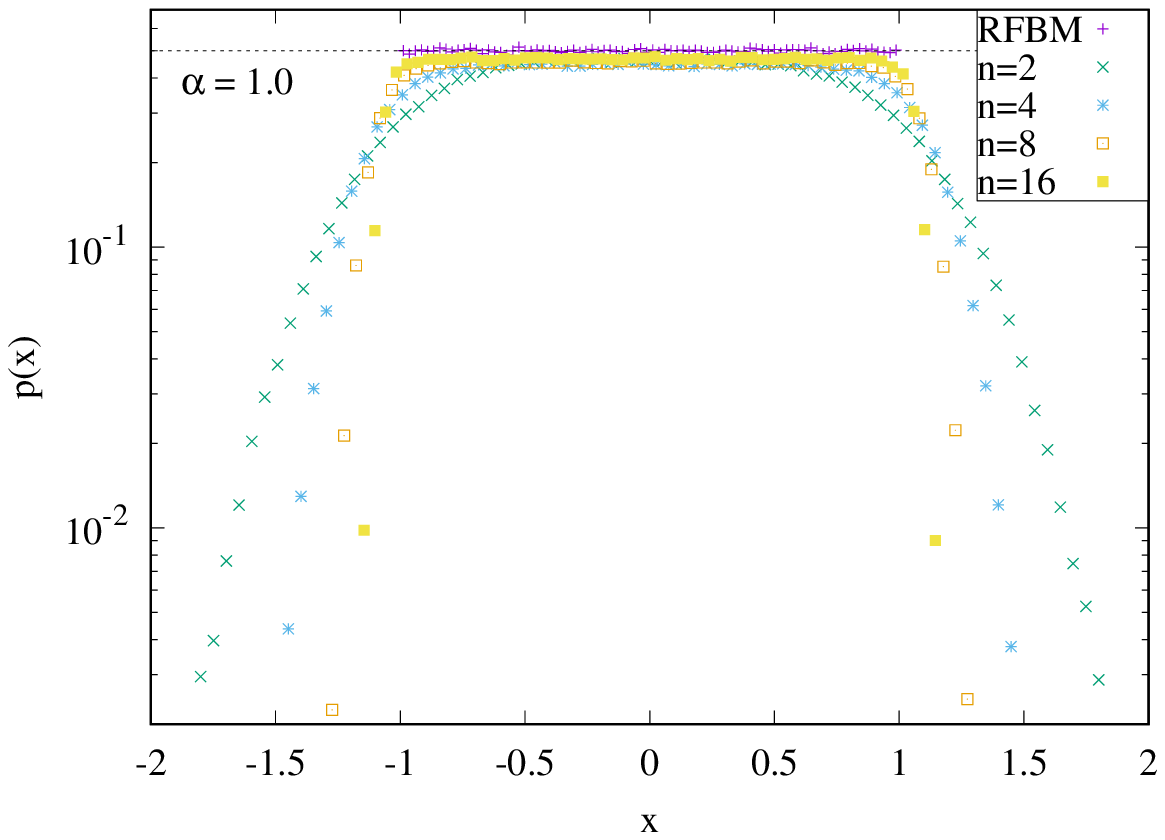}
\caption{Top left: Stationary PDF for the quartic potential with $n=2$ and
different subdiffusive exponents $\alpha<1$ (negatively correlated FGN).
The lines show fits with the function \eref{eq:supGauPdf}, where the fit
parameter was $a$. The "stretching" parameter $b$ was fixed from the
observed kurtosis value in table \ref{tab:kurtosis} by inverting the
kurtosis formula \eref{eq:curtSuperGauss} for the super-Gaussian. The case
of normal diffusion ($\alpha=1$) is shown for comparison. Top right:
stationary PDF for $\alpha=1.8$ and different superharmonic potentials.
The data labelled "RFBM" show the stationary PDF of reflected FBM in the
interval $[-1,1]$ \cite{guggenberger2019}. For increasing $n$ the potential
approaches the box shape with infinite walls, and the stationary PDF clearly
tends to the stationary PDF in the reflected case. Note that for $x<0$ the
data are plotted logarithmically (left axis) while for $x\geq0$ they are
plotted linearly (right axis). Bottom: Plots of the stationary PDF along
with the PDF for RFBM, for $\alpha=0.6$ (left) and $\alpha=1$ (right).}
\label{fig:statPdfCorrNoiseSupHarmPotFit}
\end{figure}

As for negatively correlated FGN the monomodality of the PDF is preserved
at all times but the shape is no longer Gaussian, it may be useful to have an
empirical shape for the Boltzmannian. We propose the modified Gaussian
(compare \cite{jeon2016,beta})
\begin{equation}
\label{eq:supGauPdf}
P(x)=\mathscr{N}_2\exp\left(-a[U(x)]^b\right)
\end{equation}
with the normalisation $\mathscr{N}_2$ depending on the parameters $a>0$ and $b>0$
that in turn both may depend on $n$ and $\alpha$. For
$b<1$ this would correspond to a "stretched" form as compared to the Boltzmann
distribution (\ref{eq:boltzmannForm}) for the uncorrelated case ($\alpha=1$),
for which we have $b=1$.\footnote{For positively correlated FGN ($\alpha>1$) the
stationary PDF cannot be of the shape (\ref{eq:supGauPdf}) because of the observed
bimodality.} To validate this conjecture we fitted the modified Gaussian
\eref{eq:supGauPdf} to the simulated stationary PDF in the quartic potential
($n=2$). Before fitting the parameter $b$ was taken from table \ref{tab:kurtosis}
for the corresponding kurtosis value, such that the only free parameter is $a$.
This extracted value of $b$ is also listed in table \ref{tab:kurtosis}.
The results shown in figure \ref{fig:statPdfCorrNoiseSupHarmPotFit} on the left
demonstrate that the empirical form (\ref{eq:supGauPdf}) provides a quite accurate
quantitative description of the stationary PDF. We note here that non-Gaussian
shapes can be determined not only by the kurtosis. For the analysis of data
deviations from a Gaussian can be determined from large-deviation analyses
\cite{samu} or via the codifference \cite{jakub}.

Finally, figure \ref{fig:statPdfCorrNoiseSupHarmPotFit} (right) shows the stationary
PDF for different superharmonic potentials and compares them to the stationary PDF
of reflected FBM, which was obtained numerically in \cite{guggenberger2019}.
As can be seen, the stationary PDF converges to the corresponding stationary
PDF of reflected FBM in the interval $[-1,1]$. This, of course, does not come
as a surprise since for $n\to\infty$ the superharmonic potential converges to
the infinite potential well in the interval $[-1,1]$, in which the diffusing
particle moves freely inside and is reflected at the borders. However, this
convergence adds support to the claims in \cite{guggenberger2019}: while for
reflecting boundaries one may debate the correct implementation in the
numerical scheme, for an external potential the stochastic description
(\ref{eq:langevin}) appears fully natural.

\subsection{Overdamped fractional Langevin equation motion}
\label{sec:overdampedFle}

To demonstrate that the deviations from the Boltzmannian for FBM are due to the
non-existence of the FDT, we briefly study the stationary PDF for
the overdamped FLE \cite{zwanzig2001,kou2008}. In our dimensionless variables
and for $1<\alpha<2$,
\begin{equation}
\label{eq:gle}
\alpha(\alpha-1)\int_0^t(t-t')^{\alpha-2}\frac{dX(t')}{dt'}dt'+\frac{dU}{dx}(X(t))
=\eta(t),
\end{equation}
where $\eta(t)$ is a normalised FGN and $U(x)=x^{2n}/(2n)$. 
Here the friction term includes a power-law memory that
is coupled to the noise autocorrelation via the FDT
$\langle\eta(t_1)\eta(t_2)\rangle=k(t_1-t_2)/2$, where the memory kernel is
$k(t-t')=\alpha(\alpha-1)(t-t')^{\alpha-2}$.
The integral on the left hand side of (\ref{eq:gle}) can be interpreted as a
fractional differential operator, therefore the name FLE \cite{lutz}. The
equilibrium Boltzmann distribution encoded in this equation for the superharmonic
potential $U(x)$ is given by equation \eref{eq:solStatFpeBrownian2}
with the equilibrium second moment
\begin{equation}
\label{eq:dimlessCanSecMom}
\langle X^2\rangle=\frac{n^{1/n}\Gamma(3/(2n))}{\Gamma(1/(2n))}. 
\end{equation}

For the harmonic potential ($n=1$) the overdamped FLE can be solved analytically
\cite{kou2008,kursawe2013}. The result is a Gaussian, that is fully characterised
by the first two moments
\begin{equation}
\langle X(t)\rangle=x_0E_{2-\alpha}\left(-\frac{t^{2-\alpha}}{\Gamma(1+\alpha)}
\right)
\end{equation}
and
\begin{equation}
\label{mlmom}
\langle X^2(t)\rangle=\frac{1}{2}+\left(x_0^2-\frac{1}{2}\right)E_{2-\alpha}^2
\left(-\frac{t^{2-\alpha}}{\Gamma(1+\alpha)}\right).
\end{equation}
Here we introduced the Mittag-Leffler function, that has the series expansions
around zero and infinity \cite{bateman}
\begin{equation}
E_\lambda(-z)=\sum_{\nu=0}^\infty\frac{(-1)^\nu z^\nu}{\Gamma(1+\lambda \nu)}
\sim\sum_{\nu=1}^N\frac{(-1)^{\nu-1}}{\Gamma(1-\lambda \nu)z^\nu}+O\left(\frac{1}{
z^{N+1}}\right),\,\,\, z\in\mathbb{R}^+
\end{equation}

\begin{figure}
\centering
\includegraphics[width=0.49\textwidth]{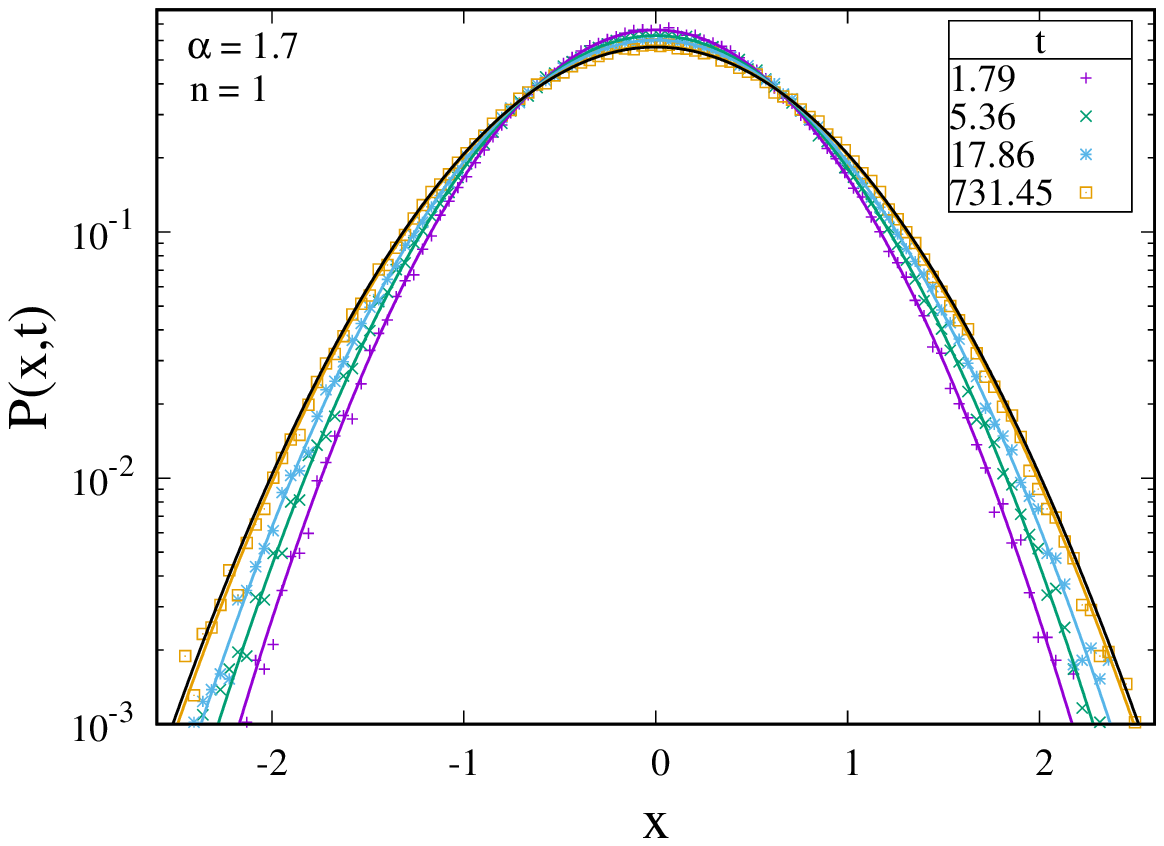}
\includegraphics[width=0.49\textwidth]{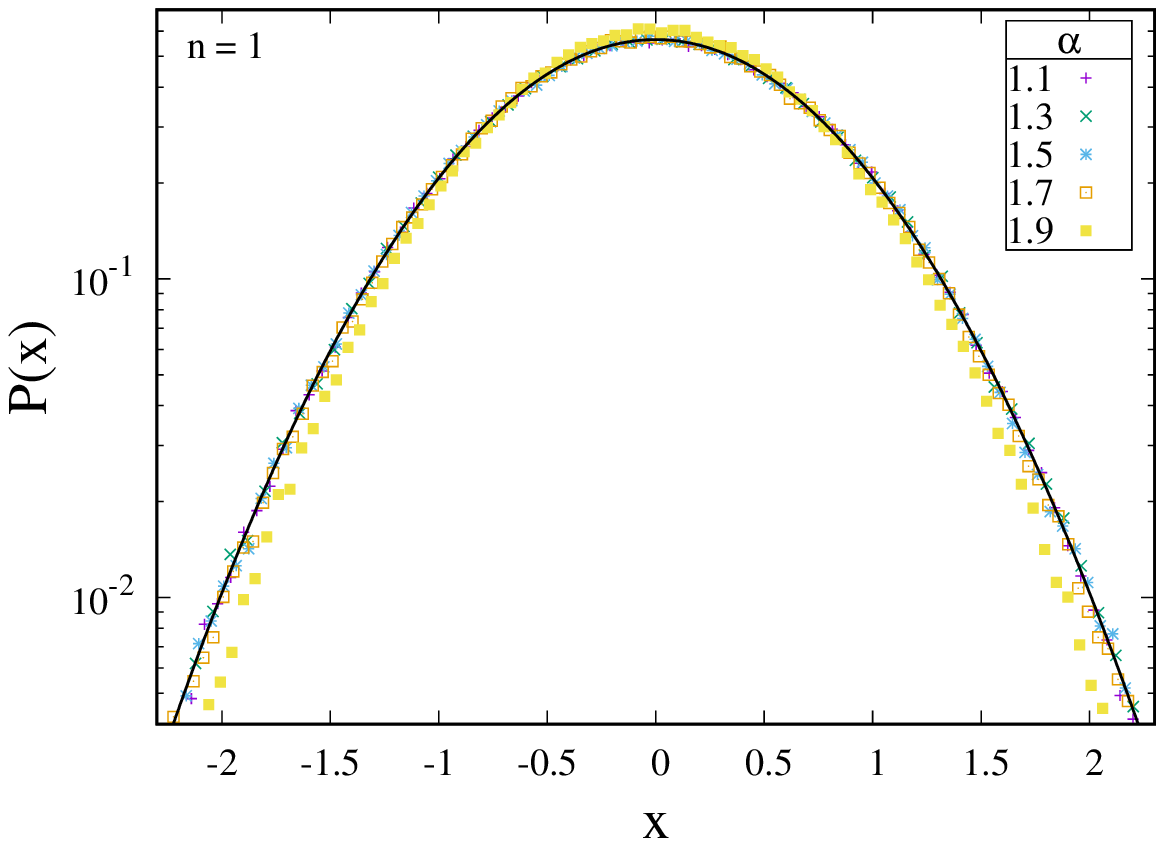}
\caption{Time-dependent (left) and stationary (right) PDF for fractional
Langevin equation motion in an harmonic potential ($n=1$), for $x_0=0$.
The coloured lines show the theoretical time-dependent PDF, a Gaussian
with the second moment \eref{mlmom}, the black line shows the stationary
PDF $P(x)=\pi^{-1/2}\exp(-x^2)$.}
\label{fig:ovdamFleHarPotPdf}
\end{figure}

In our dimensionless variables the second moment has the stationary value
$\langle X^2\rangle=1/2$, such that the stationary PDF reaches the Gaussian
Boltzmann form $P(x)=\pi^{-1/2}\exp(-x^2)$. Figure \ref{fig:ovdamFleHarPotPdf}
shows the time-dependent PDF for $\alpha=1.7$ and the stationary PDF for different
$\alpha$. Both for the time-dependent and the stationary PDFs we see very good
agreement with the predicted Gaussian forms. The slight deviation from the
theoretical stationary PDF for the two largest $\alpha$-values is due to the
fact that stationarity was not fully reached at the maximal simulation time.

\begin{figure}
\centering
\includegraphics[width=0.49\textwidth]{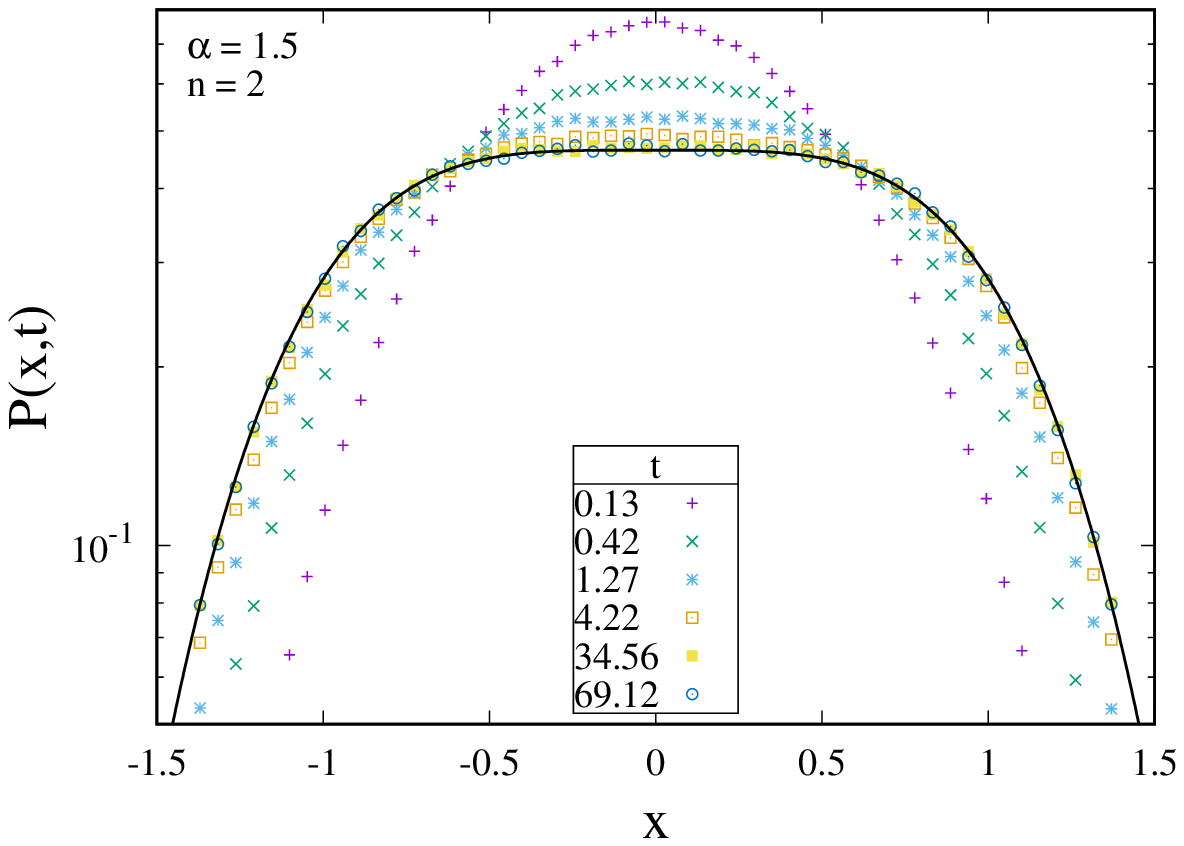}
\includegraphics[width=0.49\textwidth]{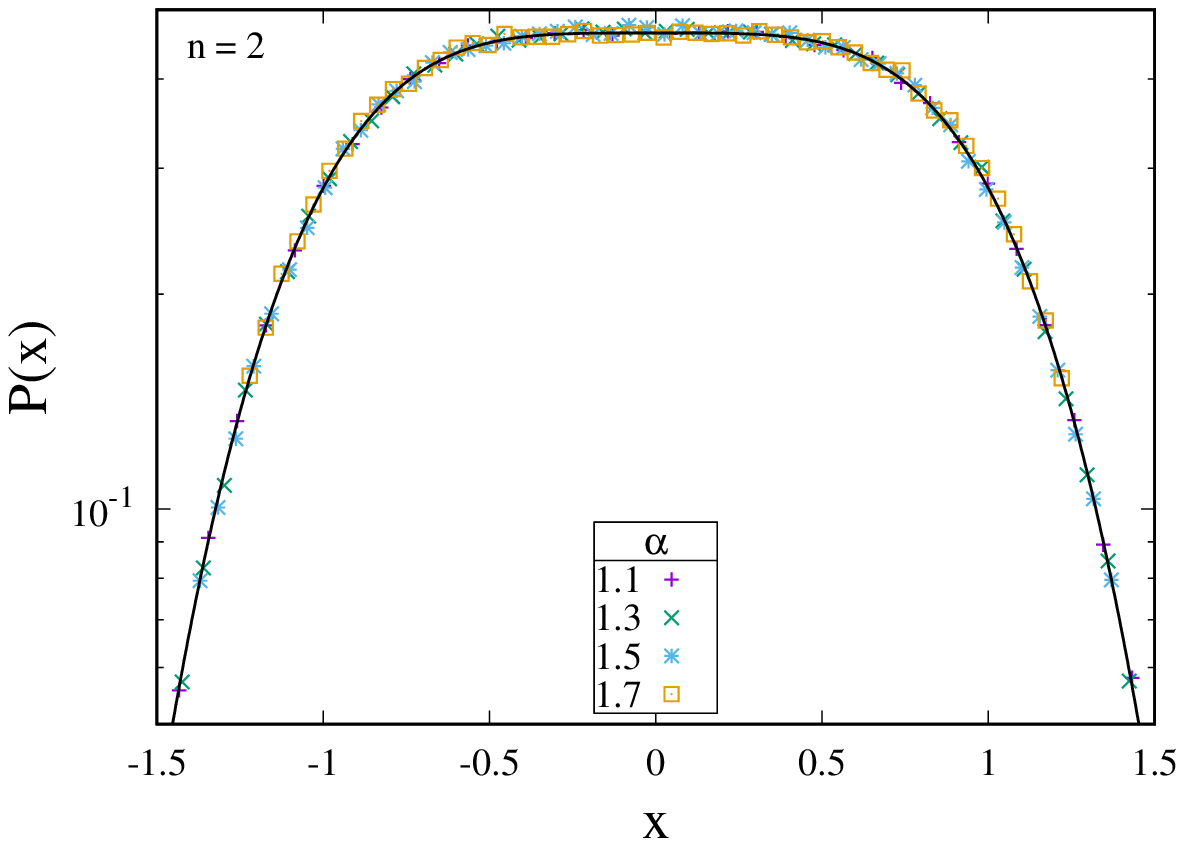}
\includegraphics[width=0.49\textwidth]{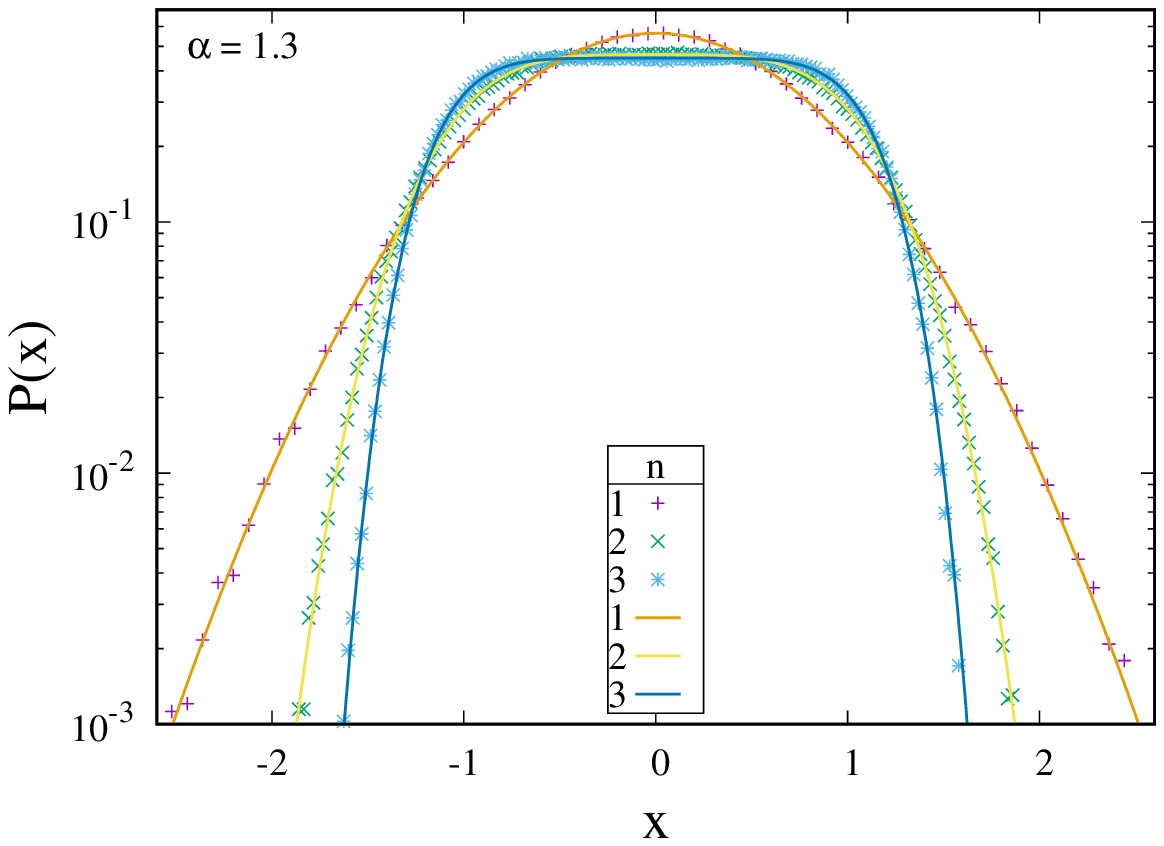}
\caption{Top: Time-dependent (left) and stationary (right) PDF for fractional
Langevin equation motion in the quartic
potential ($n=2$). Bottom: Stationary PDF for $\alpha=1.3$ and different
superharmonic potentials. The lines show the (corresponding) Boltzmann
equilibrium PDF \eref{eq:solStatFpeBrownian2}, demonstrating excellent agreement.}
\label{fig:flePdf}
\end{figure}

Figure \ref{fig:flePdf} shows the time-dependent PDF for $\alpha=1.5$ and the
stationary PDF for different $\alpha$ for the quartic potential with $n=2$ as
well as the stationary PDF for $\alpha=1.3$ and different superharmonic
potentials. The PDF stays unimodal in all cases and the stationary PDF
agrees very well with the Boltzmann distribution \eref{eq:solStatFpeBrownian2}.

\section{Conclusion}
\label{sec:conclusion}

We studied FBM in steeper than harmonic external potentials on the basis of
the overdamped Langevin equation driven by long-range correlated FGN. Our
central finding is that the stationary PDF for this
non-thermalised process significantly
deviates from the naively expected Boltzmann form. Notably, for the
positively correlated case we obtain a bimodal stationary PDF for steeper
than harmonic external potentials. The maxima were shown to be located in
close vicinity of the maximal potential curvature. As shown by the kurtosis
the obtained stationary PDF decays faster than the Boltzmannian for the
case of uncorrelated white Gaussian noise. For negatively correlated FGN the
PDF remains monomodal but the tails decay slower than those of the Boltzmann
form. In the latter case we showed that the observed stationary PDF is
consistent with the empirical form $P(x)\propto\exp(-a[U(x)]^b)$, where the
"stretching" exponent $b$ was determined from the kurtosis value. Choosing
increasingly steep external potentials we also demonstrated that the PDF
in the long time limit converges to the non-uniform shapes obtained for
reflected FBM in a box potential in \cite{guggenberger2019}.

\begin{figure}
\centering
\includegraphics[width=5.12cm]{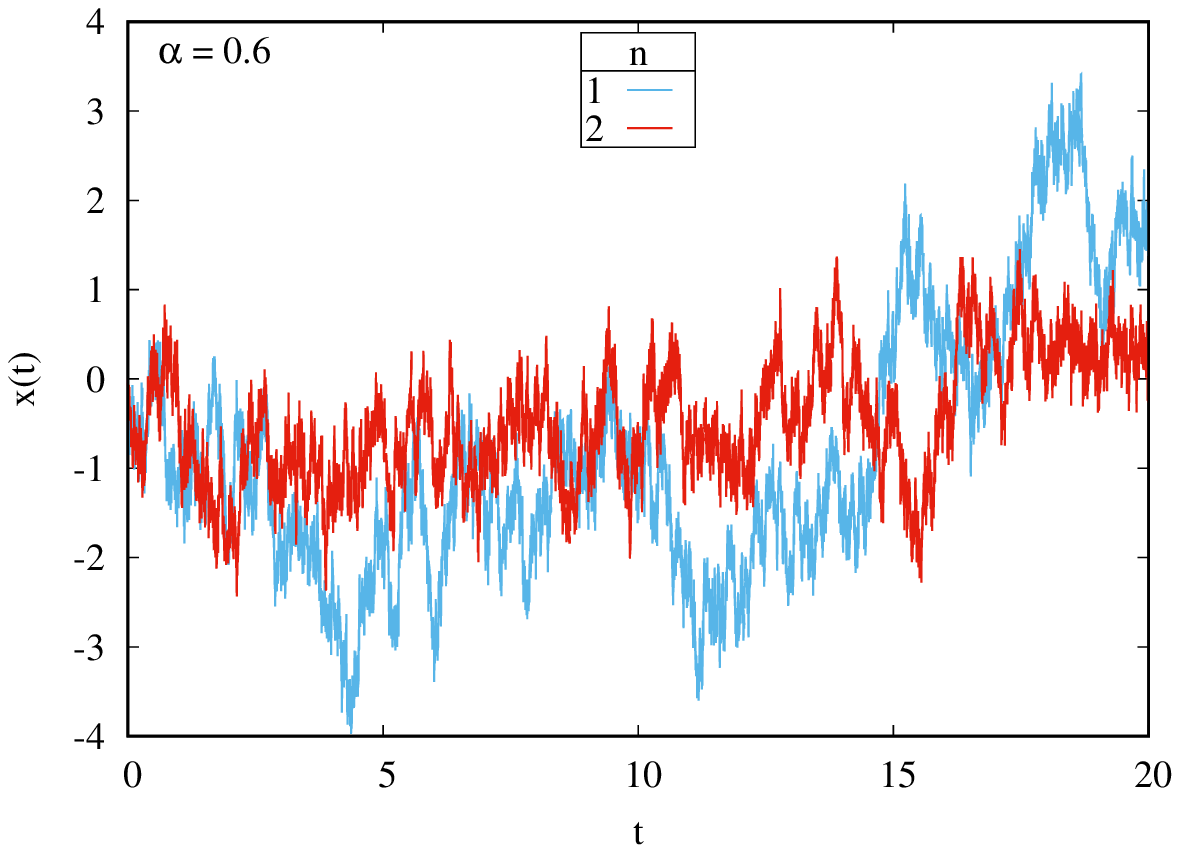}
\includegraphics[width=5.12cm]{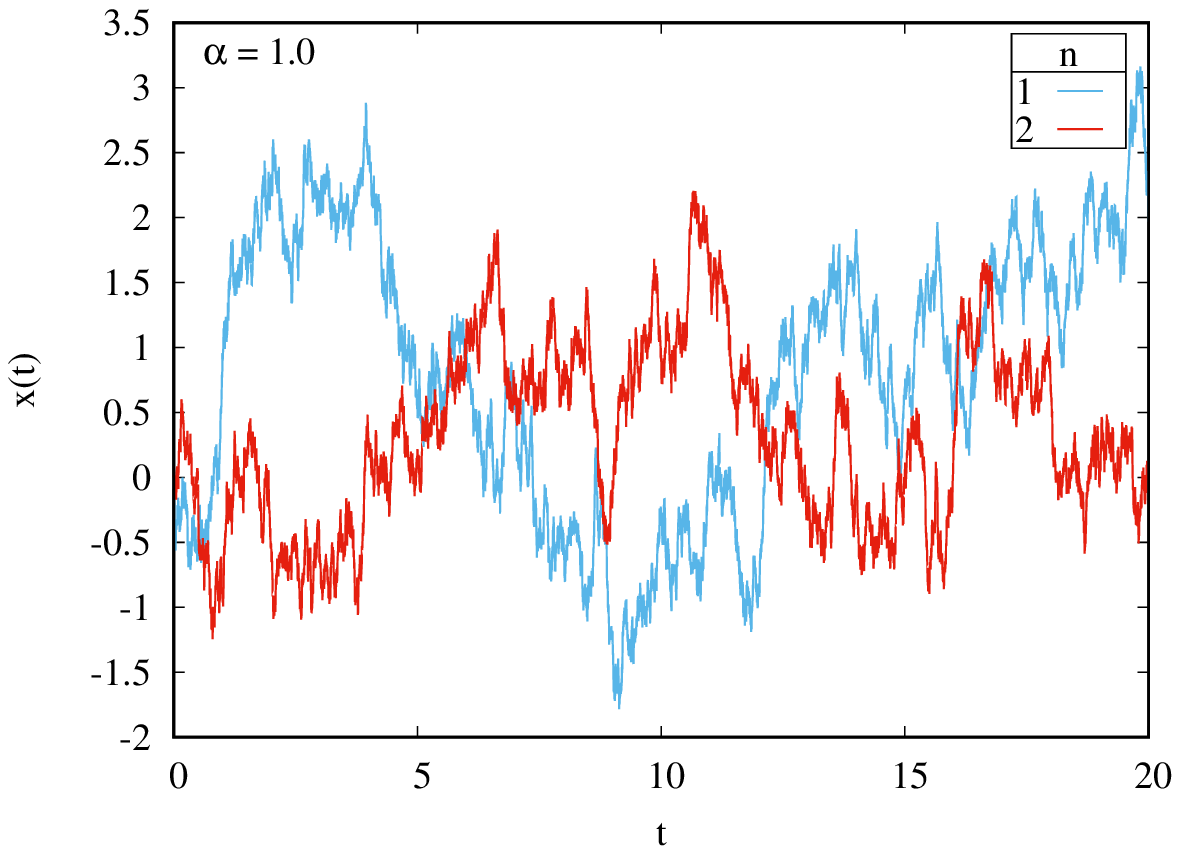}
\includegraphics[width=5.12cm]{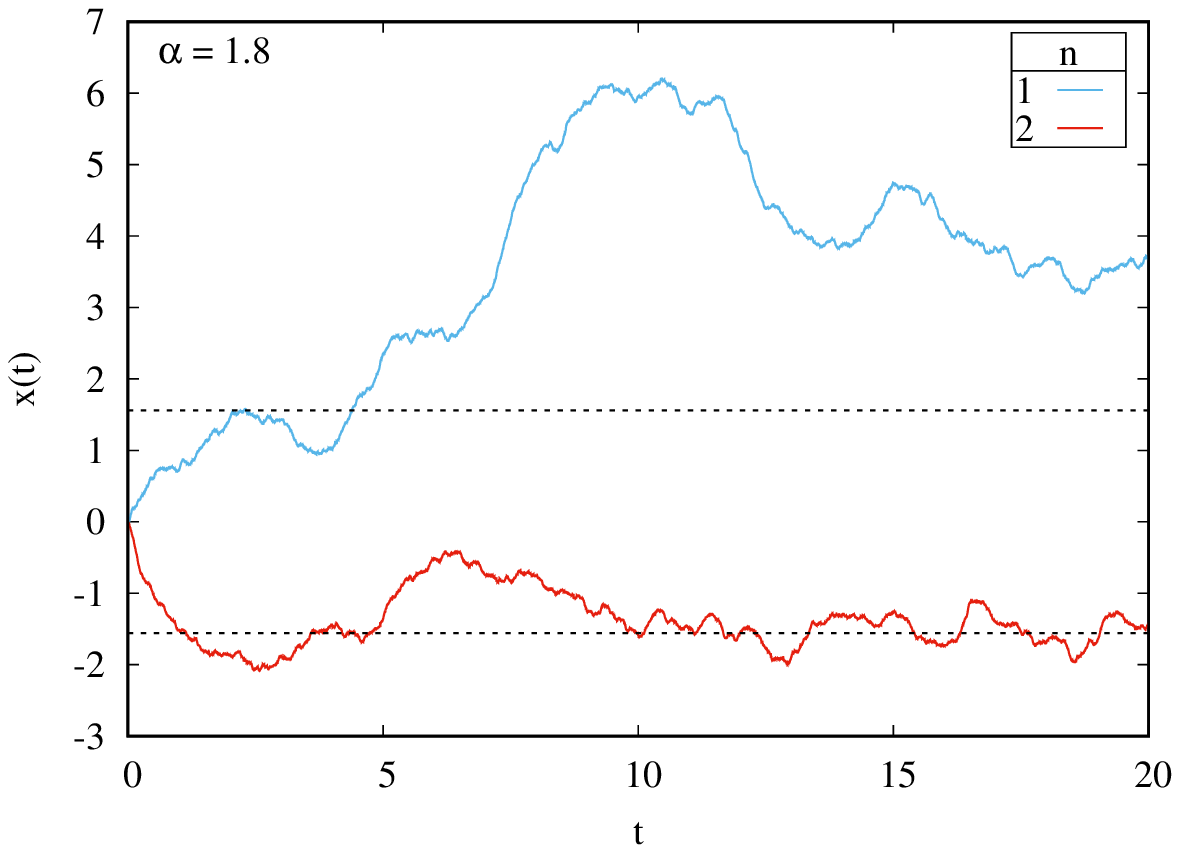}\\
\includegraphics[width=5.12cm]{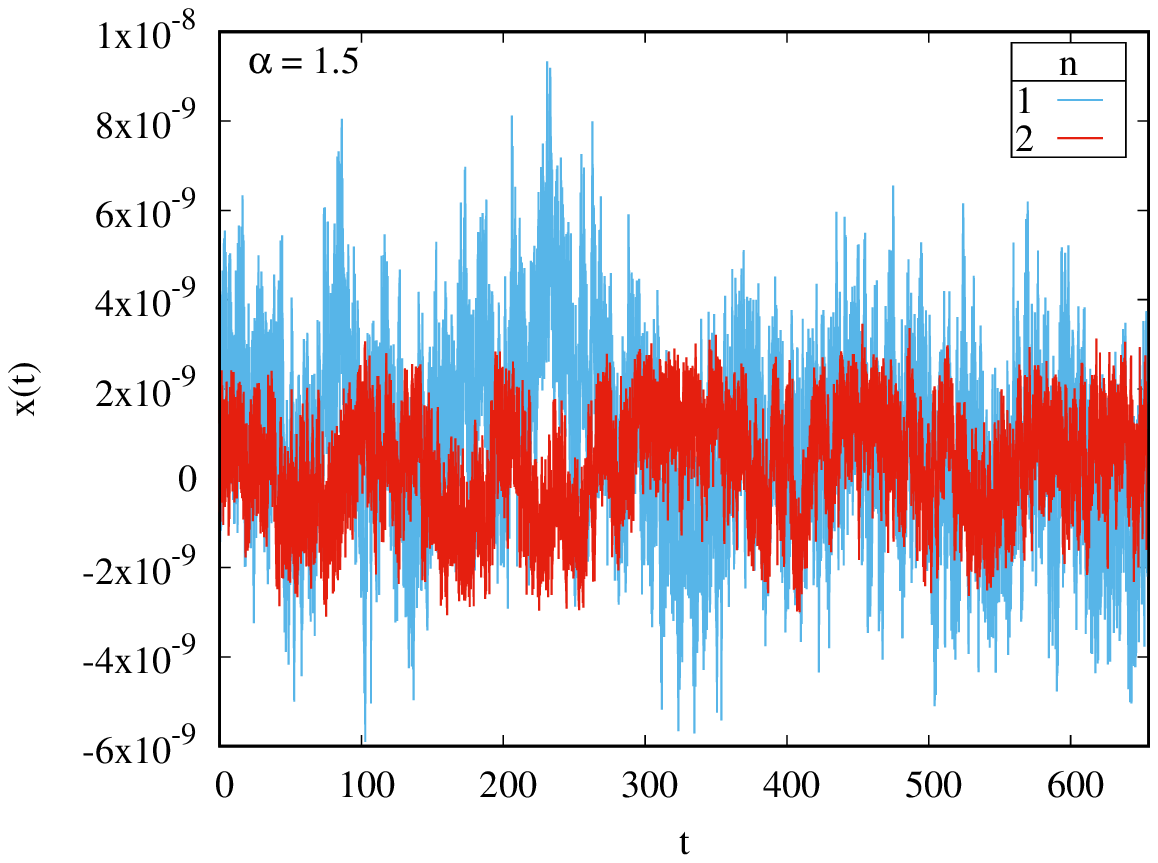}
\includegraphics[width=5.12cm]{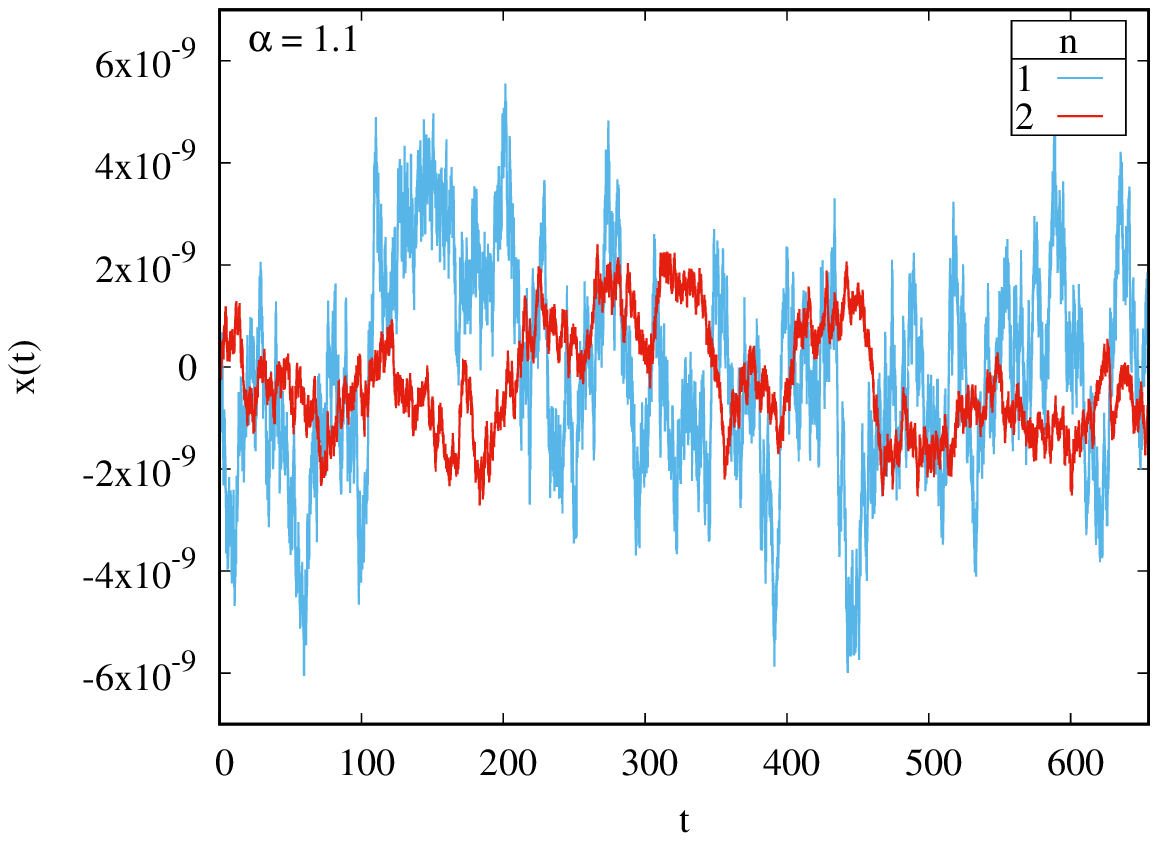}
\caption{Sample trajectories in confining potentials in
dimensional units. Top row:
FBM trajectories in harmonic ($n=1$) and quartic ($n=2$) potentials for $\alpha
=0.6$, $\alpha=1.0$, and $\alpha=1.8$ (left to right). In the panel for $\alpha
=1.8$ the dashed lines denote the location of the two maxima of the PDF for
$n=2$. Bottom row: FLE
trajectories in harmonic ($n=1$) and quartic ($n=2$) potentials for $\alpha
=1.5$ and $\alpha=1.1$. Note that for FLE motion $\alpha=1.1$ and $\alpha=1.5$
translates into subdiffusive motion with MSD exponent $0.9$ and $0.5$,
respectively.}
\label{fig_traj}
\end{figure}

For comparison we briefly consider the sample trajectories of the different
confined dynamics considered herein. Figure \ref{fig_traj} shows trajectories
for both FBM and FLE motion in harmonic ($n=1$) and quartic ($n=2$) potentials,
for different scaling exponents $\alpha$ of the FGN. For clarity we here use
dimensional units for time $t$ and position $x$, such that the stronger
confinement in the quartic potential is visible. For FBM (top row in figure
\ref{fig_traj}) the antipersistence of the motion for the subdiffusive case
$\alpha=0.6$ is in distinct contrast to the persistent behaviour for $\alpha=1.8$.
In the fomer case the particle rarely ventures deep into the potential well and
has many zero-crossings. In the latter case the particle "pushes" against
one flank of the confining potential for considerable time spans. This effect
is clearly seen in the top right panel of figure \ref{fig_traj} for the case
$n=2$, as the trajectory wiggles around one of the two dashed lines representing
the two maxima of the corresponding stationary PDF. The case of
normal diffusion with $\alpha=1.0$ is shown for comparison. FLE motion under
confinement is shown in the bottom row for two $\alpha$. We note that for FLE
the short time expansion of the MSD (\ref{mlmom}) and $x_0=0$ reads $\langle
X^2(t)\rangle\sim t^{2-\alpha}/\Gamma(1+\alpha)$. For $1<\alpha<2$ the motion
is thus subdiffusive, and due to the FDT a higher $\alpha$ value leads to
more pronounced antipersistence. As one can see, the qualitative behaviour
between the subdiffusive FBM and FLE motion dynamics is not overly distinct
from the trajectories, such that the assessment about the convergence to an
equilibrated Boltzmann distribution is better visible in the PDFs of the
processes.

We also remark that
the bimodal shape of the stationary PDF for positive correlated noise observed
here resembles the bimodal shapes for L{\'e}vy flights in superharmonic
potentials reported in \cite{chechkin2002,chechkin2003,chechkin2004,karol}.
These L{\'e}vy flights are described by a
Markovian Langevin equation driven by white yet L{\'e}vy stable noise. In this
case the stationary distribution in the harmonic case is not a Boltzmann
distribution but a L{\'e}vy stable density \cite{sune}. The bimodality of the
PDF in the case of L{\'e}vy flights emerges due to the strongly non-local jumps
in space, characterised by the divergence of the variance of the jump lengths.
As shown here, for the PDF the strong temporal persistence of superdiffusive
FBM effects a similar bimodality. We also mention that the
stationary distribution of the end point of a grafted, semi-flexible polymer
was shown to have a bimodal distribution, where the degree of bimodality can
be tuned by the stiffness of the polymer \cite{panayotis}.

For FBM in an external potential the noise is viewed as "external", and the
system is not thermalised. This approach is often used in systems such as
living biological cells, in which energy-consuming active processes drive
the environment far from equilibrium. In contrast, when the system is
thermalised and governed by a FDT the driving FGN is coupled
to the velocity of the particle by a power-law memory kernel, described by
the overdamped FLE. In this case the stationary distribution must be given
by the Boltzmann distribution, as we confirmed for steeper than harmonic
potentials.

\ack

We acknowledge funding from the German Research Foundation (DFG) through
grant number ME 1535/7-1. RM acknowledges the Foundation for Polish Science
(Fundacja na rzecz Nauki Polskiej, FNP) for support in terms of an Alexander
von Humboldt Honorary Polish Research Scholarship.

\appendix 

\section{Numerics}
\label{sec:numerics}

We here discuss the numerical schemes to compute approximate solutions of the
overdamped Langevin and the FLE discussed in the main text. The simulation of
sample trajectories requires the simulation of sample times series of FBM and
its increment process,
which is briefly discussed first. In the following we employ an equidistant
time discretisation $t_i=i\delta t$ ($i=0,\ldots,I$) with time step $\delta
t=t/I$ and
maximal simulation time $t$.

Let $B(t)$ be a (normalised) FBM, $B(t)=\int_0^t\eta(\tau)d\tau$ with (normalised)
FGN $\eta(t)$. The time-discrete FBM $B_i=B(t_i)=B(i\delta t)$ then fulfils the
recursion relation
\begin{equation}
\label{eq:fbmRecursion}
B_0=0,\,\,\,B_i=B_{i-1}+R_{i-1},
\end{equation}
where $R_i=B_{i+1}-B_i$ are the increments of FBM with time step $\delta t$. This
increment process is a centred Gaussian process with
auto-covariance function $\langle R_iR_{i+j}\rangle=(\delta t^\alpha/2)(|j+1|^\alpha
+|j-1|^\alpha-2|j|^\alpha)$. We generated time-discrete increment sample trajectories
by use of the Hosking method \cite{dieker2004}. Time-discrete FBM sample
trajectories were then generated via the recursion relation \eref{eq:fbmRecursion}.

To obtain a numerical scheme for the (dimensionless) overdamped Langevin equation
with superharmonic potential $U(x)=x^{2n}/(2n)$ we first integrate from $t_{i-1}$
to $t_i$ and obtain
\begin{equation}
\label{eq:ovdamLEqInt}
X(t_i)=X(t_{i-1})-\int_{t_{i-1}}^{t_i}X^{2n-1}(t')dt'+[B(t_i)-B(t_{i-1})].
\end{equation}
With the (left) rectangle-rule $\int_{t_{i-1}}^{t_i}f(t')dt'\approx \delta t
f(t_{i-1})$
to approximate the integral in \eref{eq:ovdamLEqInt} we find for the following
recursion relation for the numerical solution $\hat{X}_i\approx X(t_i)$,
\begin{equation}
\label{eq:ovdamLEqScheme}
\hat{X}_0=X(0),\,\,\,\hat{X}_i=\hat{X}_{i-1}-\delta t\hat{X}_{i-1}^{2n-1}+R_{i-1}.
\end{equation}

Now we turn to the (dimensionless) overdamped FLE with superharmonic potential.
We rewrite it in the integral form
\begin{equation}
\int_0^tX^{2n-1}(t')dt'=-\alpha(\alpha-1)\int_0^t(t-t')^{\alpha-2}X(t')dt'+
\alpha x_0t^{\alpha-1}+B(t).
\end{equation}
We approximate the integral on the left hand side using the composite
trapezoidal-rule
\begin{equation}
\label{eq:compositeTrapRule}
\int_0^{t_i}f(t')dt'\approx\frac{\delta t}{2}[f(0)+f(t_i)]+\delta t\sum_{j=1}^{
i-1}f(t_j).
\end{equation}
The memory integral on the right hand side is approximated by use of the formula
\cite{diethelm2002}
\begin{equation}
\label{eq:productTrapRule}
\int_0^{t_i}(t_i-t')^{\alpha-2}f(t')dt'\approx\frac{(\delta t)^{\alpha-1}}{\alpha(
\alpha-1)}\sum_{j=0}^ia_{j,i}f(t_j),
\end{equation}
where $\alpha>1$ and
\begin{equation}
\label{eq:productTrapRule2}
a_{j,i}=\left\{\begin{array}{ll}(i-1)^{\alpha}-(i-\alpha)i^{\alpha-1},&j=0\\
(i-j+1)^{\alpha}+(i-j-1)^{\alpha}-2(i-j)^{\alpha},&1\leq j\leq i-1\\
1,&j=i\end{array}\right..
\end{equation}
We then obtain the implicit recursion relation for the approximate solution
$\hat{X}_i\approx X(t_i)$,
\begin{equation}
\label{eq:ovdamFleScheme}
\hat{X}_0=x_0,\,\,\,0=\hat{X}_i^{2n-1}+a\hat{X}_i+d_i+c_i-\frac{2}{\delta t}B_i,
\end{equation}
with $a=2(\delta t)^{\alpha-2}$ and
\begin{equation}
\label{eq:ovdamFleScheme2}
d_i=2\sum_{j=1}^{i-1}\hat{X}_j^{2n-1}+a\sum_{j=1}^{i-1}a_{j,i}\hat{X}_j,\,\,\,
c_i=x_0^{2n-1}+x_0a\left(a_{0,i}-\alpha i^{\alpha-1}\right)
\end{equation}
and initial condition $x_0 = X(0)$. We solved the implicit recursion relation
\eref{eq:ovdamFleScheme} numerically by use of Newton's method. To that end we
define the function $f(y)=c_i-\frac{2}{\delta t}B_i+d_i+ay+y^{2n-1}$. We want to find
$z$ such that $f(z)=0$ (i.e., $z=\hat{X}_i$). Given an initial guess $z_0$,
Newton's method solves for $z$ recursively via $z_{k+1}=z_k-f(z_k)/f'(z_k)$
with $k=0,1,\ldots,$ and where $f'$ denotes the derivative of $f$. Hence,
\begin{equation}
\label{eq:newtonRecursion2}
z_{k+1}=z_k-\frac{c_i-\frac{2}{\delta t}B_i+d_i+az_k+z_k^{2n-1}}{a+(2n-1)z_k^{2n-2}}.
\end{equation}
To obtain an initial guess $z_0 = \hat{X}_i^{\mathrm{ini}}$, we approximate the
integral on the left-hand side of the overdamped FLE not by the composite
trapezoidal-rule \eref{eq:compositeTrapRule} but by the composite (left)
rectangle-rule
\begin{equation}
\label{eq:compositeRectangleRule}
\int_0^{t_i}f(t')dt'\approx \delta t\sum_{j=0}^{i-1} f(t_j).
\end{equation}
This leads to
\begin{equation}
\label{eq:initialGuessOvdamFleScheme}
z_0=\hat{X}_i^{\mathrm{ini}}=\frac{1}{a}\left(-x_0^{2n-1}-d_i-c_i+\frac{2}{\delta t}
B_i\right).
\end{equation}
Equations \eref{eq:ovdamFleScheme}, \eref{eq:newtonRecursion2} and
\eref{eq:initialGuessOvdamFleScheme} were used for the computation of
the sample trajectories.

\end{document}